\documentclass{aa}
\RequirePackage{aas_macros}
\usepackage[dvips]{graphicx}
\usepackage{txfonts}
\usepackage{natbib}
\bibpunct{(}{)}{;}{a}{}{,}
\usepackage{url}

\begin{document}

\title{Cosmological radiative transfer for the line-of-sight proximity effect}

\author{Adrian M. Partl\inst{1,2} \and Aldo Dall'Aglio\inst{1} \and Volker
  M\"{u}ller\inst{1} \and Gerhard Hensler\inst{2}}

\institute{Astrophysikalisches Institut Potsdam, An der Sternwarte 16, 
D-14482 Potsdam, Germany \and Institute of Astronomy, University of Vienna, 
T\"{u}rkenschanzstrasse 17, A-1180 Vienna, Austria}

\offprints{A. Partl: apartl@aip.de}

\date{Received ; Accepted}

\titlerunning{Cosmological radiative transfer for the LOS proximity effect}

\abstract{}{We study the proximity effect in the Ly$\alpha$ forest
around high redshift quasars as a function of redshift and environment employing 
a set of 3D continuum radiative transfer simulations.}
{The analysis is based on dark matter only simulations at redshifts 3, 4, 
and 4.9 and, adopting an effective equation of state for the baryonic matter, 
we infer the HI densities and temperatures in the cosmological box. 
The UV background (UVB) and additional QSO radiation with 
Lyman limit flux of $L_{\nu_{\rm LL}}=10^{31} \; \mbox{and} \; 10^{32} \mbox{erg Hz}^{-1}\mbox{s}^{-1}$
are implemented with a Monte-Carlo continuum radiative transfer code
until an equilibrium configuration is reached. We analyse 500
mock spectra originating at the QSO in the most massive halo, in a 
random filament and in a void. The proximity effect is studied using 
flux transmission statistics, in particular with the  normalised optical depth 
$\xi=\tau_{{\rm eff,\, QSO}}/\tau_{{\rm eff,\, Ly\alpha}}$, which is the ratio of the effective 
optical depth in the spectrum near the quasar and in the average  Ly$\alpha$ 
forest.}
{Beyond a radius of $r > 1 \textrm{ Mpc } h^{-1}$ from the quasar, we measure 
a transmission profile consistent with geometric dilution of the QSO ionising radiation. 
A departure from geometric dilution is only seen, when strong absorbers intervene the line-of-sight.
The cosmic density distribution around the QSO causes a large scatter in the 
normalised optical depth. The scatter decreases with increasing redshift and 
increasing QSO luminosity. The mean
proximity effect identified in the average signal over 500 lines of sight
provides an average signal that is biased by random large scale density 
enhancements on scales up to $r \approx 15 \textrm{ Mpc } h^{-1}$. 
The distribution of the proximity effect strength, a parameter which describes a shift of the 
transmission profile with respect to a fiducial profile, provides a measure of
the proximity effect along individual lines of sight. It shows a clear maximum 
almost without an environmental bias. Therefore this maximum can be used for an 
unbiased estimate of the UVB from the proximity effect.
Differing spectral energy distributions between the QSO and the UVB modify 
the profile which can be reasonably well corrected analytically. A few Lyman 
limit systems have been identified that prevent the detection of the proximity effect 
due to shadowing.}
{}


\maketitle

\section{Introduction}

The baryon content of intergalactic space is responsible for the large
number of absorption lines observed in the spectra of high redshift
quasars. This phenomenon, also known as the Ly$\alpha$ forest 
\citep{Sargent:1980fk,Rauch:1998it}, is mainly attributed to 
intervening \ion{H}{i} clouds along the line of sight (LOS) towards a QSO. 
The majority of the \ion{H}{i} Ly$\alpha$ systems are optically thin to 
ionising radiation. The gas is kept in a high ionisation state by the 
integrated emission of ultra-violet photons originating from the overall populations 
of quasars and star-forming galaxies: the ultra-violet background field 
\citep[UVB:][]{haardt96,fardal98,Haardt:2001yq}. Accurate estimates of the 
UVB intensity at the Lyman limit are crucial for the understanding of the 
relative contribution of stars and quasars to the UVB and also 
to ensure realistic inputs for numerical simulations of structure formation 
\citep{Dave:1999kk, Hoeft:2006dp}.

The intensity of the UVB can be strongly affected by energetic sources such 
as bright QSOs leading to an enhancement of UV photons in its environment. 
The neutral fraction of \ion{H}{i} consequently drops yielding a significant 
lack of absorption around the QSOs within a few Mpc. Knowing the luminosity of the QSO at 
the Lyman limit, this ``proximity effect'' has been widely used in 
estimating the intensity of the UVB \citep{Carswell:1987yq, Bajtlik:1988kx} 
at various redshifts on large samples 
\citep[][but compare \citealt{Kirkman:2008yq}]{Bajtlik:1988kx,Lu:1991rw,Scott:2000ys,Liske:2001uq,Scott:2002zl} and recently also 
towards individual lines of sight \citep{Lu:1996jt,DallAglio:2008uq}.

Principal obstacles in the analysis of the proximity effect arrise due to
several poorly understood effects which in the end might lead to 
biased estimates of the UVB. The most debated influence is the
gravitational clustering of matter around QSOs. Clustering enhances the number 
of absorption lines on scales of a few proper Mpc, 
which if not taken into account might lead to 
an overestimation of the UVB by a factor of a few. Alternatively, 
star forming galaxies in the QSO-environment may contribute to an enhancement, or
over-ionisation, of the local ionised hydrogen fraction, i.e. we would underestimate 
the UVB from the measured QSO luminosity. In the original formalism of
the proximity effect theory introduced by \citet{Bajtlik:1988kx}
(hereafter BDO), the possibility of density enhancements near the 
QSO redshift was neglected. However, \citet{loeb:1995kl} and 
\citet{Faucher-Giguere:2008gf} 
found that the UVB may be overestimated by a factor of about 3 
due to high environmental density. 
Comparing observations and simulations, \citet{DOdorico:2008qd} found large 
scale over densities over about 4 proper Mpc. The disagreement 
between the UVB obtained via the proximity effect and from flux transmission
statistics has been used to estimate the average density profile around the QSO 
\citep{Rollinde:2005uq, Guimaraes:2007uq, Kim:2008ye}. 
Only recently \citet{DallAglio:2008ua} showed that a major
reason for overestimating the proximity effect is not a general cosmic 
overdensity around QSOs,
but the methodological approach of estimating the UVB intensity by 
combining the proximity effect signal over several sight lines. Providing a 
definition of the proximity effect strength for a given LOS, they propose 
instead to investigate the strength distribution for the QSO sample, 
yielding consistency with the theoretical estimates of the UVB.

Numerical simulations of structure formation have been a crucial tool in 
understanding the nature and evolution of the Ly$\alpha$ forest. One of the 
major challenges in the current development of 3D simulations represents the 
implementation of radiative transfer into the formalism of hierarchical
structure formation \citep{Gnedin:2001fk,Maselli:2003,Razoumov:2005qf,
Rijkhorst:2006bh,Mellema:2006ve,Pawlik:2008ly}. 
In particular the coupling of  radiative transfer with the density 
evolution requires a large amount of computational resources and remains up 
to now a largely unexplored field. Therefore, most available codes apply 
radiative transfer as a post processing step.

Only a couple of radiative transfer studies exist for the Ly$\alpha$
forest \citep{Nakamoto:2001dq,Maselli:2005}. They discuss the influence
of radiative transfer on the widely used semi-analytical model of
the forest by \citet{Hui:1997fk}. \citet{Maselli:2005} 
find that due to self-shielding of the UVB flux in over-dense regions
the hydrogen photo-ionisation rate fluctuates by up to 
20 per cent and the helium rate by up to 60 per cent. 
Since radiative transfer is not negligible for a background field, 
this should be true for point sources as well. 
Up to now numerical studies dealt with ionisation
bubbles in the pre-reionisation era or right at the end of reionisation
\citep{Iliev:2007cr,Kohler:2007uq,Maselli:2007nx,Trac:2007kx,Zahn:2007fk}.
These studies aim at determining the sizes of \ion{H}{ii} regions 
as observed
in transmission gaps of Gunn-Peterson troughs \citep{Gunn:1965oq} in 
high-redshift sources, or in 21cm emission or absorption signals. 
By comparing radiative transfer simulations
with synthetic spectra, \citet{Maselli:2007nx} 
showed, that the observationally deduced size of the \ion{H}{ii} regions 
from the Gunn-Peterson trough is underestimated by up to 30 per cent.

In the study of the proximity effect at redshifts lower than that of 
reionisation, the influence of radiative transfer has been neglected
up to now, as the universe is optically thin to ionising radiation. 
We expect that absorption 
of QSO photons by intervening dense regions might reduce the QSO 
flux. These dense regions can shield themselves from the 
QSO radiation field \citep{Maselli:2005} and would not experience
the same increase in the ionisation fraction as low density regions. This 
leads to a dependence of the proximity effect on the QSO environment. 
Furthermore, 
the amount of hard photons in the QSO spectral energy distribution affects 
the proximity effect profile as suggested by \citet{DallAglio:2008uq}. 
Similar to the study of high redshift \ion{H}{ii} regions where
ionisation fronts are broadened due to the ionising flux's shape of the spectral 
energy distribution \citep{Shapiro:2004ph,Qiu:2007it}, we expect 
the size of the proximity effect zone to be a function of the spectral hardness. 
Harder UV photons have smaller ionisation cross sections and cannot 
ionise hydrogen as effectively as softer UV photons. 

%
\begin{figure*}
\begin{minipage}[t][1\totalheight]{0.49\textwidth}%
\includegraphics[bb=40bp 150bp 994bp 994bp,clip,width=1\columnwidth]
                 {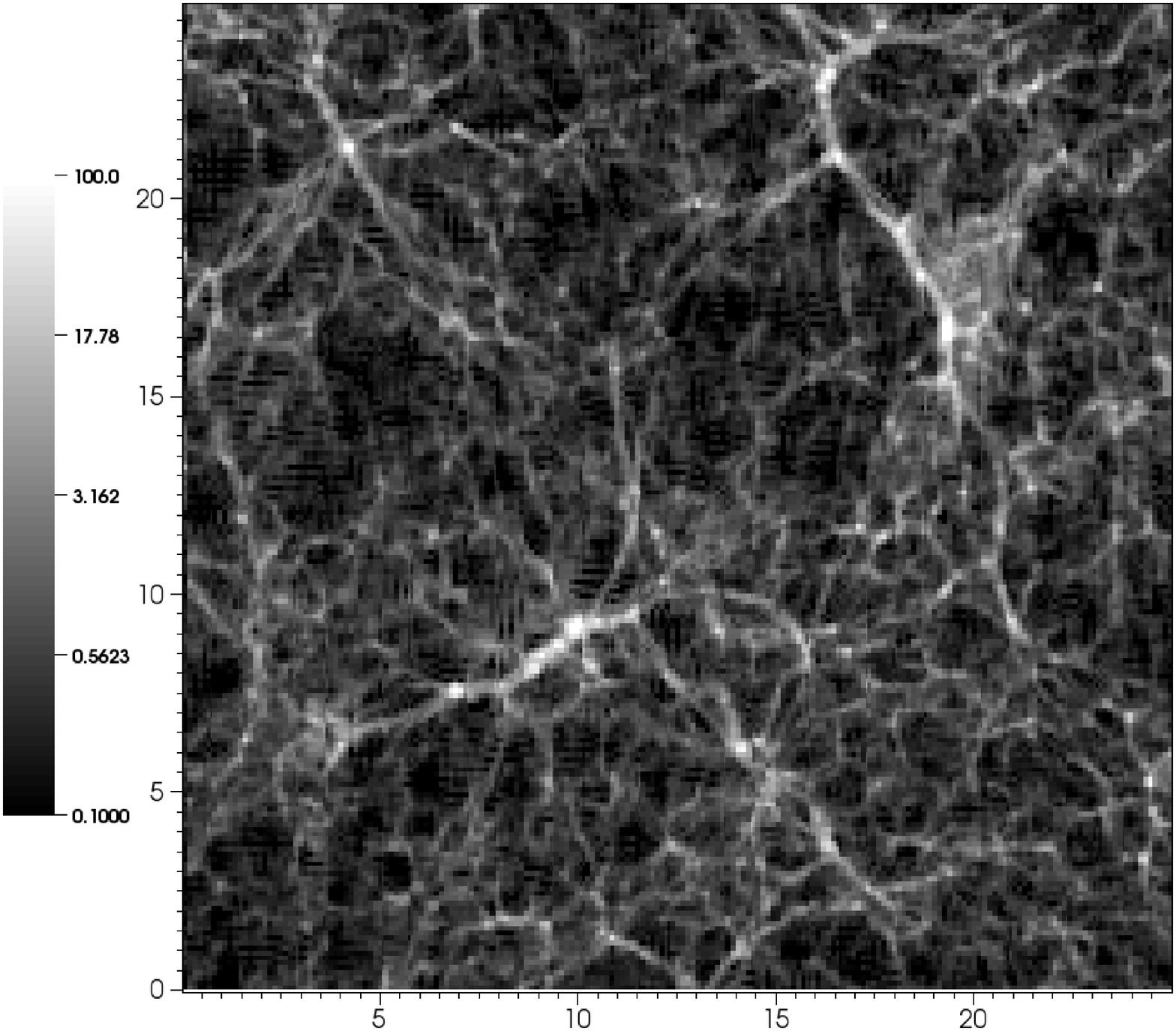}%
\end{minipage}%
\hfill{}%
\begin{minipage}[t][1\totalheight]{0.49\textwidth}%
\includegraphics[bb=40bp 150bp 994bp 994bp,clip,width=1\columnwidth]{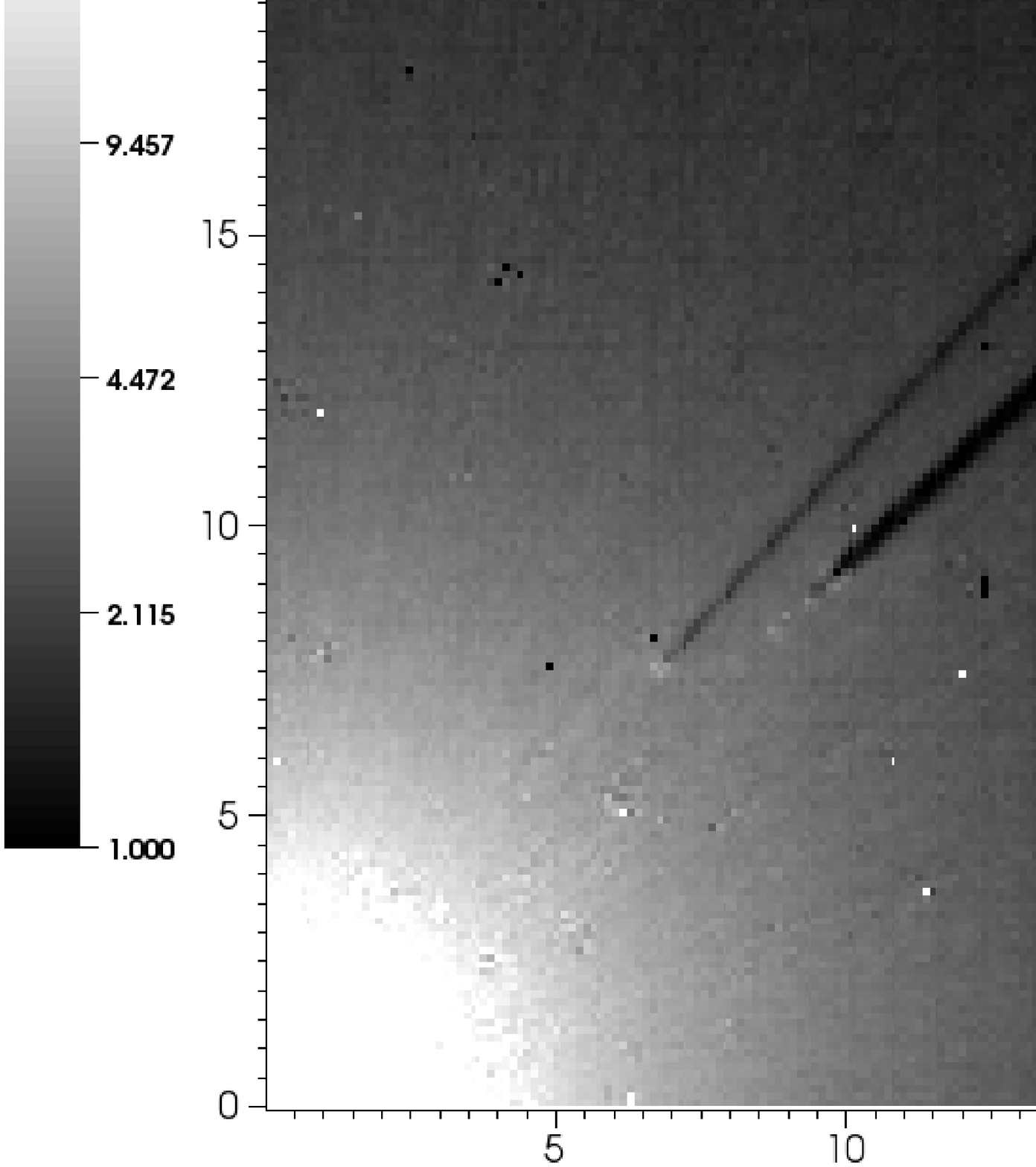}%
\end{minipage}%
\caption{\label{fig:shadowingShielding}Cuts through the $25\,\textrm{ Mpc }h^{-1}$
simulation box at redshift $z=4$ for a QSO sitting in the void at the lower
left corner. In the left
panel the over-density field $1+\delta$ is shown. The right panel provides the 
inverse of the over-ionisation fraction 
$n_{\ion{H}{i},\mathrm{Ly-\alpha}}/n_{\ion{H}{i},\mathrm{prox}}$ 
due to a 
$L_{\nu_{\rm LL}}=10^{31}\textrm{ erg Hz}^{-1}\textrm{ s}^{-1}$
QSO. Here $n_{\ion{H}{i},\mathrm{Ly-\alpha}}$ 
and $n_{\ion{H}{i},\mathrm{prox}}$ are the
neutral hydrogen fractions without and with the influence of the QSO. 
Clearly visible are the extended over-ionisation zone and shadowing
effects. The scattered white dots arise from numerical noise. The
axes are in comoving $h^{-1}\textrm{ Mpc}$.}
\end{figure*}

In this study we employ a three dimensional
radiative transfer simulation to study \ion{H}{ii} regions expanding in a
pre-ionised intergalactic medium (IGM) at redshifts $z = 3, 4, \,\mbox{and}\,
4.9$. We intend to 
quantify the influence of the above mentioned effects. To this end, we 
consider two realistic QSO luminosities. Furthermore, we study the 
influence of environment by placing the QSO either inside a  
massive halo, in a random filament, and in an under-dense 
region (void). 

The paper is structured as follows. 
In Sect. \ref{sec:Simulations} we discuss the dark matter
simulation used in this study, and we show that a realistic model of the 
Ly$\alpha$ forest is found from a semi-analytical model of the IGM.
In Sect. \ref{sec:RTInCosm} we describe the method to solve the radiative 
transfer equation, and we discuss in Sect. \ref{sec:UVB} and \ref{sec:QSO}
how the different UV sources are implemented.
In Sect. \ref{sec:modell} we 
review the standard approach used to characterise the proximity effect 
\citep{Bajtlik:1988kx}.
In Sect. \ref{sec:RTEffects} we describe the different radiative transfer
effects on the overionisation profile. Then in Sect. \ref{sec:LOSResult} we introduce
the proximity effect strength parameter and develop additional models to disentangle various
biases in the signal. In Sect. \ref{sec:meanProx}, we present the 
results for the mean line of sight proximity effect as determined from synthetic spectra.
In Sect. \ref{sec:PESD} we discuss the proximity effect strength distribution of our
spectra. Finally we summarise our findings in Sect. \ref{sec:Conclusions}.

\section{\label{sec:Simulations}Simulations}

\subsection{\label{sub:modelIGM} Initial Realisation (IGM Model)} 

\begin{figure}
\includegraphics[bb=9bps -5bps 402bps 197bps, clip, width=1\columnwidth]{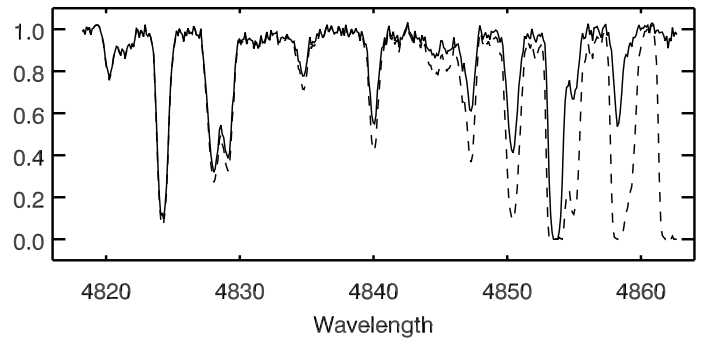}
\caption{\label{fig:sampSpec}Mock spectra synthesised from the $z=3$
snapshot with a QSO in a filament. The QSO sits on the right hand side of each spectrum, and the
wavelength is in {\AA}ngstr\"om. The dashed line gives the Ly$\alpha$ forest
spectrum without the influence of the QSO, 
the solid line the spectrum from the radiative transfer simulation
of the proximity effect for a QSO with a Lyman limit luminosity of 
$L_{\nu_{\rm LL}}=10^{31}\textrm{ erg
Hz}^{-1}\textrm{s}^{-1}$. The spectra have a resolution of $\Delta v = 6.7 \textrm{ km s}^{-1}$
and a signal to noise of 70.}
\end{figure}

We employ a DM simulation in a periodic box
 \footnote{Distances are given as comoving distances unless otherwise stated.}
of $50 \textrm{ Mpc } h^{-1}$ 
with $512^{3}$ DM particles \citep{von-Benda-Beckmann:2008fk}. 
Using the PM-Tree code GADGET2 \citep{Springel:2005dq} 
our simulations yield a force resolution of $2 \textrm{ kpc } h^{-1}$  
and a mass resolution of $m_{p}=7.75\times10^{7}\;M_{\sun}h^{-1}$.
We fix the cosmological parameters to be in agreement with the 
3rd year WMAP measurements \citep{Spergel:2006uq}:
The total mass density parameter is $\Omega_{m,0}=0.3$, 
while the baryon mass density is $\Omega_{b,0}=0.04$, and the vacuum energy is
$\Omega_{\Lambda}=0.7$. 
The dimensionless Hubble constant is $h=0.7$, and the power spectrum 
is normalised by the square root of the linear mass variance 
at $8 \textrm{ Mpc } h^{-1}$, $\sigma_{8}=0.9$.

Simulating the Ly$\alpha$ forest is a challenging task, since a large
box is required to capture the largest modes that still influence 
the mean flux and its distribution \citep{Tytler:2009sp, Lidz:2010wt}. 
Furthermore high resolution is required to capture the properties of single
absorbers correctly. Our simulations yield a reasonable representation
of the average statistics of the Lyman alpha forest 
\citep{Theuns:1998rw, Zhang:1998oz, McDonald:2001xr, Viel:2002fe} for low redshifts. 
However recently it has been shown by \citet{Bolton:2009th} and \citet{Lidz:2010wt}
that to resolve the Ly$\alpha$ forest properly at high redshifts, the mass resolution has to 
be increased by two orders of magnitudes for $z=5$ and one order of magnitude
for $z=4$. Given the large box sizes needed, these resolution constraints cannot be fully
met with todays simulations. According to \citet{Bolton:2009th}, the error in our estimate 
of the mean flux at $z=4$ is around 10\%. For $z=3$, the simulations will be marginally converged
with mass resolution.

We record the state of the simulation at three different 
redshifts $z=4.9,\, 4,\, \textrm{and }3$. Using cloud in cell 
assignment we convert the DM particle distribution into a density and
velocity field on a $400^3$ regularly spaced grid.

We then select three different environments from the highest redshift snapshot: 
The most massive halo with a mass of $M_{\rm{halo}} = (2.6
\times 10^{12}, 4.0 \times 10^{12}, 7.9 \times 10^{12}, 1.0 \times 10^{15}) 
M_{\sun} h^{-1}$ at redshifts $z=(4.9, 4, 3, 0)$, a random filament, and a random void. 
The last two environments are selected by visual 
inspection of the particle distribution and their position is tracked down to $z=3$. 
These locations will be considered to host a QSO.

To better characterise these environments, we estimate the volume-averaged
overdensities $\delta_5$ in $r=5\textrm{ Mpc } h^{-1}$ spheres and find 
$\delta_5=(1.7, 1.8, 2.1)$ at redshifts $z=(4.9, 4.0, 3.0)$ 
for the halo, $\delta_5=(1.0, 0.8, 1.1)$
for the filament, and $\delta_5=(0.8, 0.7, 0.7)$ for the void.
Thus the halo resembles a locally overdense region, the void an locally underdense
region, and the filament an average environment.
Figure~\ref{fig:shadowingShielding} illustrates a snapshot of the DM density 
field at redshift $z=4$.

In order to describe the baryonic component of the IGM, we assume a
universe containing hydrogen only, whose density and velocity fields are
proportional to those of the dark matter \citep{Petitjean:1995yq}.
In order to account for pressure effects on baryons, 
\citet{Hui:1997fk} proposed to convolve the density field with a window function
cutting off power below the Jeans length.
In \citet{Viel:2002vf} differences of the gas and DM densities between hydrodynamical and 
DM only simulations are studied. For $\delta \approx 3$, baryons follow the DM distribution quite
well, i.e. over most of the densities relevant for the Lyman-$\alpha$ forest. However
at higher DM densities, the corresponding gas density is lower due to the smoothing
induced by gas pressure. We implicitly smooth our density field with the cloud in cell density
assignment scheme and cell sizes of 125 $h^{-1}$ kpc. This is comparable to the Jeans length
of $\approx 150 \, h^{-1} \textrm{ kpc}$ at $z=3$ and mean density, which scales as $\delta^{-1/2}
(1+z)^{-3/2}$.

The hydrogen density at position $x$ is then given by 
\begin{equation}
n_{{\rm H}}(x)=\frac{3\, H_{0}^{2}\, \Omega_{b,0}}{8\, \pi\,  G\,  m_{{\rm p}}} (1+z)^{3} \left(1+\delta(x) \right) 
\label{eq:hydDens}
\end{equation}
where $G$ is the gravitational constant, 
$H_{0}$ the Hubble constant, and $m_{{\rm p}}$ the proton mass. The 
DM overdensity is $1+\delta(x)=\rho(x)/\overline{\rho}$, 
with DM density $\rho(x)$. 
Finally the hydrogen velocity field $v(x)$ is assumed to be
equal to the DM one.

\subsection{\label{sub:calib} Model and calibration of the intergalactic medium}

The thermal evolution of the IGM is mainly determined by the equilibrium of 
photoionisation heating and adiabatic cooling, resulting in a tight
relation between the density and temperature of the cosmic gas. This relation is known as 
the effective equation of state \citep{Hui:1997uq}, and is typically
expressed by $T=T_{{\rm 0}}(1+\delta)^{\gamma-1}$.
We apply the effective equation of state as an estimate of the baryonic density 
for overdensities $0.1 < \delta < 10$ relevant for the Lyman alpha forest. 
Larger densities concern collapsed regions which we approximate by assuming a cut-off 
temperature $T_{\mathrm{cut-off}} = T(\delta = 10)$.
Following the formalism of 
\citet{Hui:1997fk}, we can compute an \ion{H}{i} Ly$\alpha$ absorption 
spectrum from the density and velocity fields, once
$T_{{\rm 0}}$, $\gamma$, and the UV background photoionisation rate ($\Gamma_{\rm UVB}$) 
are fixed. Thus 
through a match between simulated and observed \ion{H}{i} absorption 
spectra properties, we can constrain these free parameters. Such a calibration
will be crucial in the following to provide realistic representations of the IGM for the
radiative transfer calculations.

\begin{figure}
\includegraphics[width=1\columnwidth]{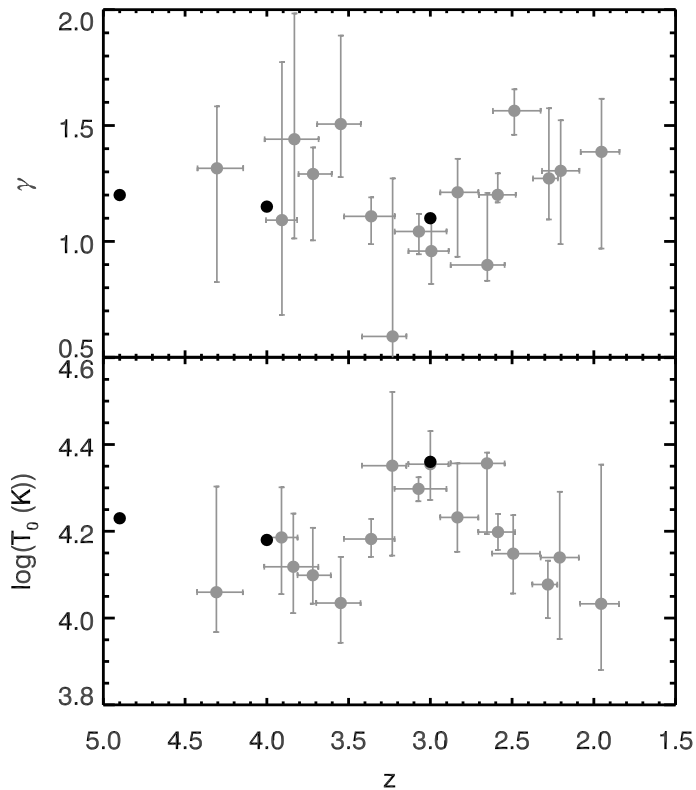}

\caption{\label{fig:EOS}\emph{Upper panel: } Comparison of our choices for $\gamma$ (black
points) with observationally derived results by \citet{Schaye:2000kx} (gray points). 
\emph{Lower panel: }Comparison of our model $T0$ (black points) with observations by
\citet{Schaye:2000kx} (gray points).}

\end{figure}

To calibrate our spectra, we employ three observational constraints sorted by
increasing importance: (i) The observed equation of state \citep{Ricotti:2000nx,
Schaye:2000kx}, (ii) the evolution of the UV background photoionisation rate 
\citep{Haardt:2001yq,Bianchi:2001ys}, and (iii) the observed evolution of the effective 
optical depth in the Ly$\alpha$ forest \citep{Schaye:2003kx, Kim:2007kx}.

Our catalogues of synthetic spectra consist of 500 lines of sight at each
redshift randomly drawn through the cosmological box. The spectra are
binned to the typical resolution of UVES spectra of $6.7 \textrm{ km s}^{-1}$
and are further convolved with the UVES instrument profile. Random noise
with a signal to noise (S/N) ratio 70 is added, the mean S/N in the QSO sample used
by \citet{DallAglio:2008ua}.
An example of such a mock spectra is presented in Fig.~\ref{fig:sampSpec}.

The model parameters of the effective equation of state $T_{{\rm 0}}$ and $\gamma$
were chosen according to observations of \citet{Ricotti:2000nx} and \citet{Schaye:2000kx}.
To obtain values for $z>4$, the observed $T_{{\rm 0}}$ and $\gamma$ parameters
have been extrapolated to higher redshifts (see Fig. \ref{fig:EOS}). 
We further constrained our model to yield an observed 
average effective optical depth $\tau_{\rm{eff}}(z) =-\ln\left\langle F(z)\right\rangle $ 
where $F$ is the transmitted flux, and the averaging is performed over the whole line of sight.
For this we used recent observations from \citet{Kim:2007kx}. With these constraints we
determined the UVB photoionisation rate $\Gamma_{\rm UVB}$ for our models.

\begin{figure}
\includegraphics[bb=0bp 0bp 360bp 278bp,clip,width=1\columnwidth]{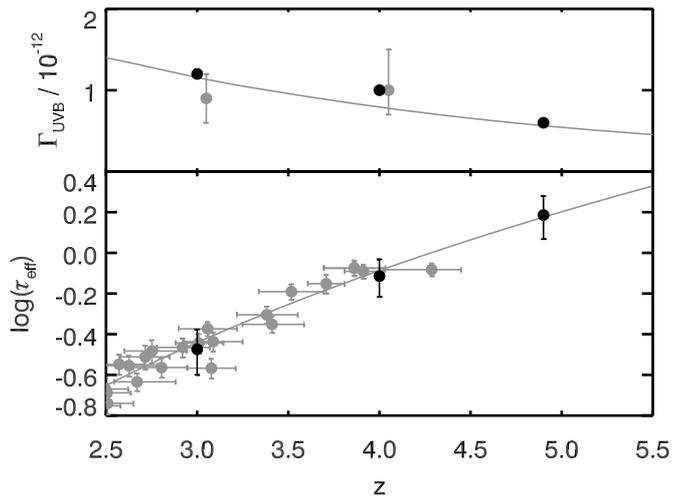}

\caption{\label{fig:meanTauEff}\emph{Upper panel: } The evolution of the UV background
photoionisation rate in our three snapshots (black points) compared to \citet{Bolton:2005yq} (grey points
shifted by $z=0.05$ for better visibility)
and predictions by \citet{Haardt:2001yq} (grey line).
\emph{Lower panel: }The effective optical depth of our models (black
points) in comparison to measurements by \citet{Schaye:2003kx} (grey
points). The continuous line shows the fit by \citet{Kim:2007kx}.}

\end{figure}

Our model parameters are presented in Table \ref{tab:gnedModels}, and
are plotted in Fig.~\ref{fig:meanTauEff} in comparison with
different literature results.
Both the inferred evolution of the UV background and the effective optical depth
closely follow recent results by \citet{Haardt:2001yq} and \citet{Bolton:2005yq}, and high
resolution observations by \citet{Schaye:2003kx} and \citet{Kim:2007kx}, respectively.

\begin{table}
\centering
\caption{\label{tab:gnedModels}Model parameters of the semi-analytical model.}

\begin{tabular}{cccccc}
\hline 
\hline$z$ & $\log T_{0}$ & $\gamma$ & $\Gamma_{\rm{UVB}}$ 
& $\log \tau_{\rm{eff}} $\\
\hline
$4.9$ & $4.23$ & $1.20$ & $0.6\times10^{-12}$ & $0.186$\\
$4.0$ & $4.18$ & $1.15$ & $1.0\times10^{-12}$ & $-0.114$\\
$3.0$ & $4.36$ & $1.10$ & $1.2\times10^{-12}$ & $-0.474$\\
\hline
\end{tabular}
\end{table}

\section{\label{sec:RTInCosm}Radiative Transfer in Cosmological Simulations}

\subsection{\label{sec:RTMethod}Method}

We now focus on the method implemented to solve the radiative
transfer equation. In developing our Monte-Carlo radiative transfer code (A-CRASH), 
we closely follow the approach first introduced by \citet{Ciardi:2001} 
and further extended by \citet{Maselli:2003}. While their latest development
includes multi frequency photon packages \citep{Maselli:2008sp},
we only implement the earlier formulation which uses monochromatic 
packages\footnote{Our A-CRASH code is OpenMP
parallel and publicly available under the GPL license at 
\url{http://sourceforge.net/projects/acrash/}}. 
The advantage of the Monte-Carlo scheme is that it accounts for the emission
of diffuse recombination photons, typically neglected in codes relying on the
``on-the-spot'' approximation (see Section \ref{sec:Diffusion} for further details).

The idea behind a Monte-Carlo radiative transfer scheme is to bundle
the radiation flux field into a discrete amount of radiation energy in form of
photon packages having
\begin{equation}
\Delta E_{i}=\int_{t_{i-1}}^{t_{i}}L_{\mathrm{UV}}(t)dt\label{eq:energyPackage}
\end{equation}
as energy content, sent out by a source with UV luminosity $L_{\mathrm{UV}}$ between two time steps 
$t_{i-1}$ and $t_{i}$. At each time step $t_{i}$ we allow the corresponding  content of energy 
$\Delta E_{i}$ to be evenly distributed
among the number of photon packages emitted by the source. In our implementation of the
Monte-Carlo scheme this number can be larger than one.
Additionally each photon is characterised by a given frequency $\nu$ following
a given spectral energy distribution (SED).
Thus, knowing the source SED and its UV luminosity we are able to infer
the number of photons $N_{\gamma, \;i}$ per package. To guarantee a proper
angular and spatial sampling of the radiative transfer equation, the photon
packages are emitted in random directions.

Once all the photon packages are produced by the sources, they are propagated through the 
computational domain. Every time a package crosses a cell, a certain amount of photons is
absorbed. The absorption probability in the $l$-th cell is 
\begin{equation}
P(\tau_{l})=1-e^{-\tau_{l}}\label{eq:absProb}
\end{equation}
with $\tau_{l}$ being the optical depth in the $l$-th cell
\begin{equation}
\tau_{l}=\sigma_{{\mbox{\scriptsize\ion{H}{i}}}}
               (\nu)\;n_{{\mbox{\scriptsize\ion{H}{i}}}, \;l}\;f_{l}\Delta x
\label{eq:localTau}
\end{equation}
where $\sigma_{{\mbox{\scriptsize\ion{H}{i}}}}$ is the hydrogen 
photo-ionisation cross-section, $n_{{\mbox{\scriptsize\ion{H}{i}}}, \;l}$ 
the neutral hydrogen number density, and $f_{l}\Delta x$ the crossing path 
length of a ray through a cell with size $\Delta x$. 
We calculate the exact crossing length using the
fast voxel traversal algorithm by \citet{amanatides:1987}.

The number of absorbed photons in the cell is 
$N_{A, \;l}=N_{\gamma, \;l}\;P(\tau_{l})$
where $N_{\gamma, \;l}$ denotes the remaining number of photons in the package 
$N_{\gamma, \;l}=N_{\gamma, \;l-1}-N_{A, \;l-1}$ arriving at the $l$-th cell. 
In our simple case of pure hydrogen gas, we can write 
\begin{equation}
n_{{\rm H}}\frac{dx_{{\mbox{\scriptsize\ion{H}{ii}}}}}{dt}=
\gamma_{{\mbox{\scriptsize\ion{H}{i}}}}(T)n_{{\mbox{\scriptsize\ion{H}{i}}}}n_{{\rm e}}-
\alpha_{{\mbox{\scriptsize\ion{H}{ii}}}}(T)n_{{\mbox{\scriptsize\ion{H}{ii}}}}n_{{\rm e}}+
\Gamma_{{\mbox{\scriptsize\ion{H}{i}}}}n_{{\mbox{\scriptsize\ion{H}{i}}}}\label{eq:chemEqu}
\end{equation}
where $x_{{\mbox{\scriptsize\ion{H}{ii}}}}=
n_{{\mbox{\scriptsize\ion{H}{ii}}}}/n_{{\rm H}}$ is the ionisation fraction,
$n_{{\rm H}}$ 
is the total hydrogen density, $\gamma_{{\mbox{\scriptsize\ion{H}{i}}}}(T)$
is the collisional ionisation rate,
$\alpha_{{\mbox{\scriptsize\ion{H}{ii}}}}(T)$ 
the recombination rate, and $\Gamma_{{\mbox{\scriptsize\ion{H}{i}}}}$ is the 
photoionisation rate derived from the number of absorbed photons $N_{A, \;l}$. 
We adopt $\gamma_{{\mbox{\scriptsize\ion{H}{i}}}}(T)$ and $\alpha_{{\mbox{\scriptsize\ion{H}{ii}}}}(T)$
as in \citet{Maselli:2003}. We solve this stiff differential equation using a 4th order Runge-Kutta scheme.

The extent of a time step is defined
as the total simulation time $t_{s}$ divided by the total number of photon packages $N_{p}$ 
emitted by each source. The numerical resolution of
the simulation is determined by calculating the mean number of packages
crossing a cell 
\begin{equation}
N_{{\rm cr}}=\frac{N_{s}N_{p}}{N_{c}^{2}}\gg\frac{t_{s}}
      {t_{{\rm min}}}\label{eq:numCross}
\end{equation}
where $N_{{\rm s}}$ is the number of sources and $N_{{\rm c}}$ the
number of cells in one box dimension, and $t_{{\rm min}}$ is the smallest characteristic
time scale of all the processes involved. In order to efficiently parallelise
the scheme, the original rule of producing one photon per source per
time step is dropped; a source is allowed to produce more
than one package per time step as long as $\Delta t\ll t_{{\rm min}}$.

We have confirmed that our implementation passes the simple tests
described by \citet{Maselli:2003} (see \citet{Partl:2007gf} for further details). 

\subsection{\label{sec:UVB}UV Background Field}

 The UV background photoionisation rates given in Table \ref{tab:gnedModels} need to be modelled in 
the framework of the radiative transfer scheme. We assume the spectral
shape of the UVB to be a power law $\nu^{\alpha_\mathrm{b}}$
with $\alpha_\mathrm{b} = -1.3$ \citep{Hui:1997fk} which is a bit lower than
recent measurements by \citet{Fechner:2006si} yielding $\alpha_\mathrm{b} = -1.99 \pm 0.34$. 
From the spectral energy distribution of the UV background field and its
intensity, photon packages need to be constructed. To obtain the number of
photons in a background package, the total energy
content of the background field in the box is mapped to single photon
packages; we follow the method by \citet{Maselli:2005}, but for the hydrogen 
only case.

To derive the energy content of a single background field photon package $N_{\gamma}$, 
the total amount of energy carried by the background field in the whole simulation box needs
to be considered. The change in the mean ionised 
hydrogen density $n_{{\mbox{\scriptsize\ion{H}{ii}}}}$ by the UVB is 
\begin{equation}
\Delta n_{{\mbox{\scriptsize\ion{H}{ii}}}}=
 \Gamma_{\rm UVB}n_{{\mbox{\scriptsize\ion{H}{i}}}}\Delta t,\label{eq:ionEquiCell}
\end{equation}
where $\Gamma_{\rm UVB}$ is the photo-ionisation rate of the background field.
This is compared to the total amount of absorbed photons in the box $N_{\gamma} \;(1-\exp(-\Delta\tau))$,
where $\Delta\tau = \sigma_{{\mbox{\scriptsize\ion{H}{i}}}} \; \Delta n_{{\mbox{\scriptsize\ion{H}{i}}}} \; d_{{\rm B}}$
is the total optical depth in the box of length $d_{{\rm B}}$. This corresponds to a mean change of the
ionised hydrogen number density of
\begin{equation}
\Delta n_{{\mbox{\scriptsize\ion{H}{ii}}}}=\frac{N_{\gamma}\left(1-e^{-\Delta\tau}\right)}
     {d_{{\rm B}}^{3}}. \label{eq:photoAbsInCell}
\end{equation}
By equating Eqs. \ref{eq:ionEquiCell} and \ref{eq:photoAbsInCell} and assuming an optically thin medium
$\Delta\tau\ll1$ with $\Delta \tau\approx\left(1-e^{-\Delta\tau}\right)$ we obtain the photon number content
of a background photon package
\begin{equation}
N_{\gamma}=\frac{\Gamma_{\rm UVB}\Delta t\left(\Delta x\cdot
    N_{c}\right)^{2}}{\sigma_{{\mbox{\scriptsize\ion{H}{i}}}}}.\label{eq:backPhotCont}
\end{equation}
Note that this is only true, if the photon package is propagated over the distance of 
exactly one box length.

Background photons are emitted isotropically from random cells
in the box. Dense regions are allowed to shield themselves from the background flux. Thus we 
emit background
photons only from cells below a certain density threshold $\delta_{{\rm UV}}$. In \citet{Maselli:2005}
a threshold of $\delta_{{\rm UV}}=60$ was used, corresponding to the density at the virial radius
of collapsed haloes. We chose a lower threshold  $\delta_{{\rm UV}}=1$ in order to ensure that mildly 
overdense regions have the possibility to shield themselves from the UV background.

\subsection{\label{sec:QSO}QSO radiation}

As with the UV background, discrete point sources are characterised by a SED and
luminosity. We only include one source of radiation other than the UVB.
The point source representing the QSO was chosen to follow the composite QSO
spectra obtained by \citet{2007AJ....133.1780T}.
The mean SED was constructed from over 3000 spectra
available in both the Galaxy Evolution Explorer (GALEX) Data Release
1 and the Sloan Digital Sky Survey (SDSS) Data Release 3. It covers
a wide wavelength range of about $9000 > \lambda\; > 300\textrm{\AA}$.

\begin{figure}
\includegraphics[bb=15bp 0bp 422bp 280bp,clip,width=1\columnwidth]{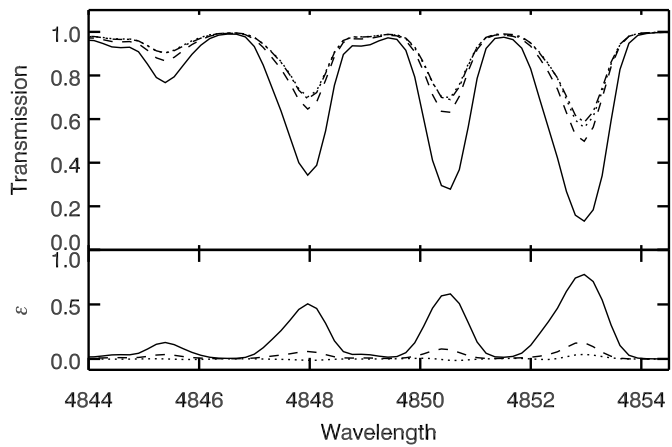}
\caption{\label{fig:convStudy} Ly$\alpha$ absorption along a sight line 
near a $z=3$ $L_{\nu_{\rm LL}}=
10^{32}\textrm{ erg Hz}^{-1}\textrm{ s}^{-1}$} QSO simulated with 
photon numbers $N_{p,QSO}=1\times10^{8}$ 
(solid line), $1\times10^{9}$ (dashed line), $2\times10^{9}$ 
(dotted line), and $4\times10^{9}$ (dash dotted line). The wavelength is in
{\AA}ngstr\"om and no noise is added to the spectra for better visibility. 
The lower panel shows the relative deviation to the highest 
resolution run.
\end{figure}

If a power law $\nu^{\alpha_\mathrm{q}}$ is assumed for the UV part of the QSO
spectrum, the data of \citet{2007AJ....133.1780T} leads to $\alpha_\mathrm{q} \approx -2.5$ 
for $\lambda < 912 \textrm{\AA}$. Since the scatter in the data of \citet{2007AJ....133.1780T} is
large in the wavelength interval we are interested in, and other authors find
harder spectra \citep{Telfer:2002yq, Scott:2004rt}, we consider 
in addition a power law SED with a 
shallower slope of $\alpha_\mathrm{q}=-1.5$. The upper energy limit at 
$\lambda = 300\textrm{\AA}$ results in a underestimation of the hydrogen photoionisation
rate of 0.6\% for $\alpha = -1.5$ and only 0.2\% for $\alpha = -2.5$ with respect to
spectra extending to higher energies.

The two QSO Lyman limit luminosities $L_{\nu_{\rm LL}}$ 
studied are chosen to bracket the luminosity range
of observed QSOs \citep{Scott:2000ys,Rollinde:2005uq,Guimaraes:2007uq,DallAglio:2008ua}.
We therefore chose the QSOs to have Lyman limit luminosities of 
$10^{31}\textrm{ erg Hz}^{-1}\textrm{ s}^{-1}$ and 
$10^{32}\textrm{ erg Hz}^{-1}\textrm{ s}^{-1}$.
In order to obtain the total UV luminosity for use with Eq. \ref{eq:energyPackage}, the high energy
part of the SED below $912\textrm{\AA}$ is scaled to the given Lyman
limit luminosity and integrated from $912>\lambda \;>300\textrm{\AA}$.
For the \citet{2007AJ....133.1780T} spectral template and
$L_{\nu_{\rm LL}}=
10^{31}\textrm{ erg Hz}^{-1}\textrm{ s}^{-1}$, 
we obtain a total UV luminosity of 
$L_{\rm UV}=1.15\times 10^{46}\textrm{ erg }\textrm{s}^{-1}$, 
and for $L_{\nu_{\rm LL}}=
10^{32}\textrm{ erg Hz}^{-1}\textrm{ s}^{-1}$
a luminosity of $L_{\rm UV}=1.15\times 10^{47}\textrm{erg}\textrm{ s}^{-1}$.

\subsection{\label{sec:SimConverge}Simulation Setup and Convergence}

Due to the high precision required to resolve the slight changes in the
ionisation fractions governing the proximity effect, a large amount of photon packages 
had to be followed. To achieve this, sub-boxes of $25\textrm{ Mpc }h^{-1}$ with $200^3$
cells have been cut out of the whole box. To ensure the resulting proximity effect region to be 
traced to as large
a radius as possible, the QSO source was located at the box origin at the cost of analysing only
$1/8$ of the full sphere around the QSO. 

A convergence study is carried out for one of our models with 
resulting mock spectra shown in Fig. \ref{fig:convStudy}. We test QSO samplings  
$N_{p,{\rm QSO}}=10^{8},\;10^{9},\:2\times10^{9},\;4\times10^{9}$ for a 
$L_{\nu_{\rm LL}}=
10^{32}\textrm{ erg Hz}^{-1}\textrm{ s}^{-1}$ QSO residing in a filament at $z=3$. For this test
the $\alpha = -2.5$ spectrum has been used. Simulating the bright QSO at $z=3$ provides the most
challenging convergence test. The IGM is already highly ionised, and the influence of the 
quasar result in very small neutral hydrogen fractions and thus need high numerical resolution. 
The lower luminosity QSO will not alter the IGM as strongly and therefore resolving the  
ionisation fractions numerically is easier. The same is true at higher redshift, where the 
IGM is not yet as strongly ionised
as at the low redshifts, and the resulting neutral fraction are larger
due to the larger optical depth. The $\alpha = -2.5$ spectrum is used for this test, since it is more
challenging to properly sample the high energy tail compared to the harder $\alpha = -1.5$ model.

It is obvious, that $10^{8}$ photons for the QSO are insufficient,
even if the mean number of photons crossing each cell is around 10. This number has 
been considered sufficient to resolve Str\"{o}mgren spheres in a homogeneous 
medium \citep{Maselli:2003}. The solutions for 
$N_{p,{\rm QSO}}=2\times10^{9}$ and $4\times10^{9}$ are similar, with relative deviations to the 
highest resolution run of at most 5 per cent. 
In order to achieve good angular sampling of the QSO environment and sufficiently large 
cell crossing numbers we consider the solution for $N_{p,{\rm QSO}}=2\times10^{9}$ as 
converged. To correctly sample the whole sphere, $1.6\times10^{10}$
packages would be required, which is at the moment beyond the capabilities
of our code. 

The Monte-Carlo radiative transfer method is able to follow photo-ionisation
heating, however since this
effect is already included in the equation of state, we keep the temperatures derived from the
equation of state fixed 
throughout the simulation.
This approach neglects any temperature fluctuations due to \ion{He}{ii} photo-heating
during \ion{He}{ii} reionisation around $z \approx 3$ \citep{McQuinn:2009by, Meiksin:2010fq}.
The simulations are run up to $t_{s}=2.5\times10^{8}\:\textrm{yr}$
and the ionisation fractions were averaged over different outputs
at different time steps to reduce the Monte-Carlo noise.

\section{\label{sec:modell}Line-of-Sight Proximity Effect}

\begin{figure*}

\begin{minipage}[t][1\totalheight]{0.48\textwidth}
\includegraphics[bb=10bp 0bp 412bp 278bp,clip,width=1\columnwidth]
     {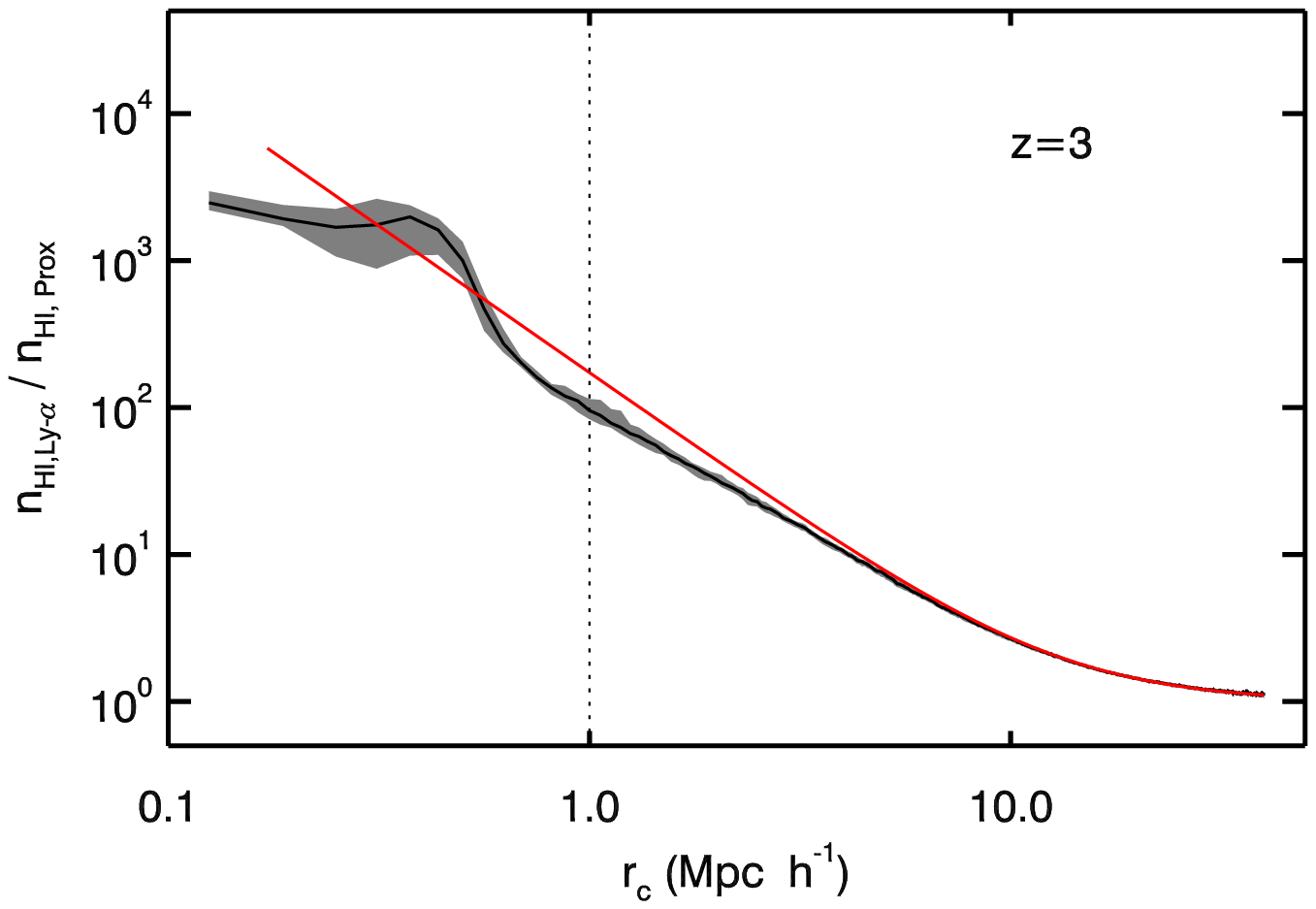}
\end{minipage}
\hfill{}
\begin{minipage}[t][1\totalheight]{0.48\textwidth}
\includegraphics[bb=10bp 0bp 412bp 278bp,clip,width=1\columnwidth]
     {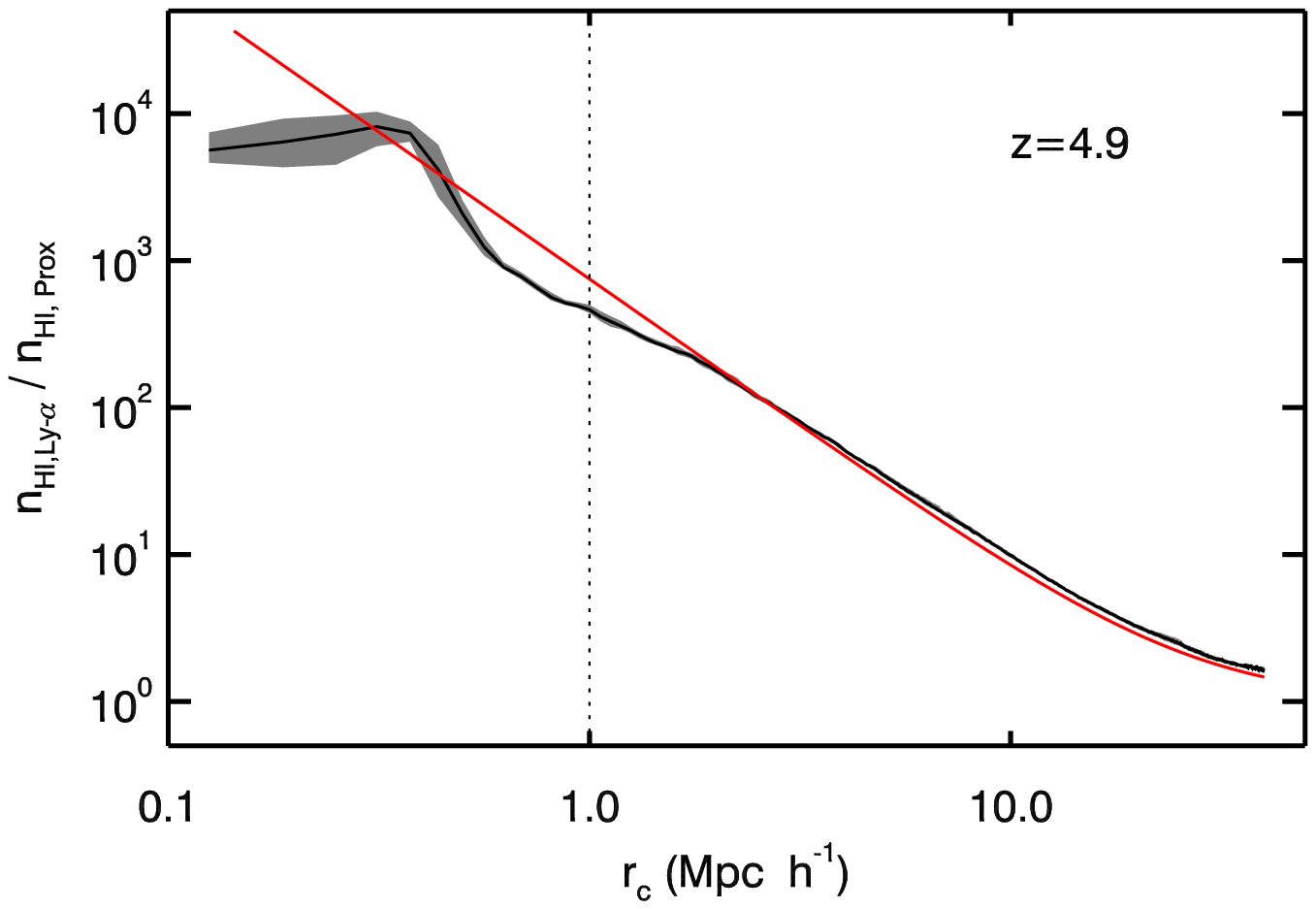}
\end{minipage}

\caption{\label{fig:ProxyProZ3} Median over-ionisation profile   
as a function of the distance from the QSO (black line, grey area gives
upper and lower quartiles) hosted by a 
halo at $z=3.0$ (left) and $z=4.9$ (right side) for a QSO luminosity 
$L_{\nu_{\rm LL}}=1\times10^{31}
\textrm{erg Hz}^{-1}\textrm{ s}^{-1}$. 
The analytical over-ionisation profile including the SED effects is given by the solid line
without the grey area. The region affected by oversampling ($r < 1 \textrm{ Mpc}$ ) is 
marked with a dotted line.}
\end{figure*}

In addition to the radiative transfer method for simulating 
the proximity effect, we also exploit a standard approach to account 
for the QSO radiation field. This method builds on the work of 
\citet[][also BDO]{Bajtlik:1988kx} and \citet{Liske:2001uq} and we briefly 
summarise it here. Under the assumption that the IGM is in photoionisation 
equilibrium with the global UV radiation field, the amount of UV photons 
increases in the vicinity of bright QSOs and dominates over the UVB. 
This effect leads to a reduction in neutral hydrogen, and with it to an
opacity deficit of the IGM within a few Mpc from the QSO.

The influence of the QSO onto the surrounding medium leads to an 
effective optical depth in the Ly$\alpha$ forest 
\begin{equation}
\tau_\mathrm{eff,QSO}(z)= \tau_\mathrm{eff, Ly\alpha}(z) \; (1+\omega(z))^{1-\beta}
\label{eq:tauProfile}
\end{equation}
\citep{Liske:2001uq} where $\tau_\mathrm{eff, Ly\alpha}(z)$ represents the evolution of the
effective optical depth in the Ly$\alpha$ forest with redshift, and $\tau_\mathrm{eff,QSO}(z)$ is the 
effective optical depth including
the alterations by the QSO radiation. Further
 $\beta$ is the slope in the column density
distribution and $\omega(z)$ is the ratio between the QSO and the UVB 
photoionisation rates.

Then following the assumption of pure geometrical dilution of the QSO
radiation as proposed by \citet{Bajtlik:1988kx}, we obtain
\begin{equation}
\omega_\mathrm{BDO}(z) = \frac{L_{\nu_{{\rm LL}}}}
  {16\pi^2 J_{\nu_{\rm LL}}(1+z) d_{L}(z_{\mathrm{q}},z)^{2}}
  \label{eq:omega}
\end{equation}
with $z$ as the redshift of absorbers along the LOS such that 
$z<z_{\mathrm{q}}$, and $d_{L}(z_{\mathrm{q}},z)$ is the luminosity distance 
from the absorber to the QSO. $L_{\nu_{{\rm LL}}}$ is the Lyman limit 
luminosity of the QSO and $J_{\nu_{\rm LL}}$ the UVB flux at the Lyman limit.
In observations, the emission redshift of the QSO $z_{\mathrm{q}}$ is subject to 
uncertainties which propagates through to the $\omega$ scale 
and increase the uncertainties in the determined UVB photoionisation 
rate \citep{Scott:2000ys}. 
In this work we consider the emission redshift to be perfectly known. 

We note that this formula is derived assuming identical spectral indices of the
QSO and the UVB. 
In our simulation we consider two different SEDs for the UVB and the 
QSO. Therefore in using Eq.~\ref{eq:omega} to estimate the proximity 
effect, we would introduce a bias. In $\omega(z)$ the two different spectral shapes can be
accounted for by the ratio of the photo-ionisation rates of the QSO as a 
function of radius $\Gamma_{\rm{QSO}}(r)$ and the photo-ionisation rate of the
UVB, $\Gamma_{\rm{UVB}}$
\begin{equation}
\omega(z)=\frac{\Gamma_{\rm{QSO}}(r)}{\Gamma_{\rm{UVB}}}=
\omega_{\rm BDO}(z)\ \frac{3-\alpha_\mathrm{b}}{3-\alpha_\mathrm{q}}.
  \label{eq:omega2}
\end{equation}
where $\alpha_\mathrm{b}$ and $\alpha_\mathrm{q}$ describe the UVB
and the QSO SED, respectively, by power laws $f_\nu \propto \nu^\alpha$. Further
it is assumed that $\sigma_{{\mbox{\scriptsize\ion{H}{i}}}} \propto \nu^{-3}$. 
Finally, the $\omega$ scale is uniquely defined once the QSO Lyman limit flux, 
redshift, and the UVB flux are known. In the following analysis we will always use Eq. \ref{eq:omega2} 
unless otherwise stated.

This correction for differing SEDs has recently been used in measurements of 
the UVB photoionisation rate \citep{DallAglio:2009ud}. 
However depending on the spectral shape of the QSO, 
omitting this correction
can result in over- or underestimation of the background flux. As stated in Section \ref{sec:QSO}
we use a QSO spectrum with a rather steep slope of $\alpha_\mathrm{q} = -2.5$. Just using the
original BDO formulation would result in an overestimation of the UV background flux by almost
30\% assuming $\alpha_\mathrm{b} = -1.3$. We will later check if this correction is able to 
model the SED effect by comparing 
our simulations with one using a shallower QSO spectral slope of 
$\alpha_\mathrm{q} = -1.5$. 

As a side note, this spectral effect will alter the QSO's sphere of
influence. We take as size of the proximity effect the
radius $r_{\omega=1}$ where \mbox{$\Gamma_{\rm{QSO}} = \Gamma_{\rm{UVB}}$}. 
Then $r_{\omega=1} \propto r_{\rm{BDO},\omega=1}
\sqrt{(3-\alpha_\mathrm{b})/(3-\alpha_\mathrm{q})}$, 
and the radius is in the case of our steep QSO spectra 13\% smaller for equal
$L_{\nu_{{\rm LL}}}$.

\section{\label{sec:RTEffects}Radiative Transfer Effects}

\subsection{\label{sec:OIPinSimulation} The Over-ionisation Profile}

The over-ionisation profile of the proximity effect around the QSO decays
radially due to geometric dilution. 
Additionally, differences in the spectral
energy distribution between the UVB and the QSO influences the size of the proximity effect zone.
Both factors are included in the $\omega$ scale. 
To determine how much the QSO decreases its surrounding neutral hydrogen
density in our simulations we determine the over-ionisation profile
$\Xi(r) = n_{\ion{H}{i},\mathrm{Ly-\alpha}} / n_{\ion{H}{i},\mathrm{Prox}}$, where 
$n_{\ion{H}{i},\mathrm{Ly-\alpha}}$ gives the neutral hydrogen fraction
unaffected by the QSO radiation. 
Further $n_{\ion{H}{i},\mathrm{Prox}}$ denotes the neutral hydrogen density 
in case of including the additional QSO radiation. Assuming ionisation equilibrium, 
the over-ionisation  profile is directly proportional to the $\omega$ scale.

In Fig. \ref{fig:ProxyProZ3} we show the median over-ionisation profile determined on 200 lines
of sight. These have been extracted from the simulation of a 
$L_{\nu_{\rm LL}}=10^{31}\textrm{ erg Hz}^{-1}\textrm{ s}^{-1}$ QSO residing in the
most massive halo at $z=3$ and $z=4.9$. For both redshifts the simulation data closely 
follows the analytical model described in Section \ref{sec:modell} at 
radii $r \gtrapprox 5\textrm{ Mpc } h^{-1}$ at $z=3$ and $r \gtrapprox 2 \textrm{ Mpc } h^{-1}$
at $z=4.9$. However, the faint QSO, the simulated over-ionisation profile starts to 
deviate at smaller radii, where the near host environment leaves an imprint in the profile. 
In contrast, for the case of the bright QSO shown in the upper left panel of Fig. \ref{fig:scatterComp}, 
the profile follows better the analytical model.
We need to emphasise that the spatial resolution of the simulations
are limited to $r = 0.125 \textrm{ Mpc } h^{-1}$. Therefore the immediate vicinity of the
QSO is badly resolved. Furthermore, cells up to 1 Mpc are subject to oversampling in all our cases.
Only beyond 1 Mpc cells are at most sampled once every time step. Therefore
the solution in this part of the over-ionisation profile, which is marked in our plots,
is unrealiable.
                                        
\subsection{Shadowing by Lyman Limit Systems}

\begin{figure}
\includegraphics[bb=9bps -5bps 402bps 197bps, clip, width=1\columnwidth]{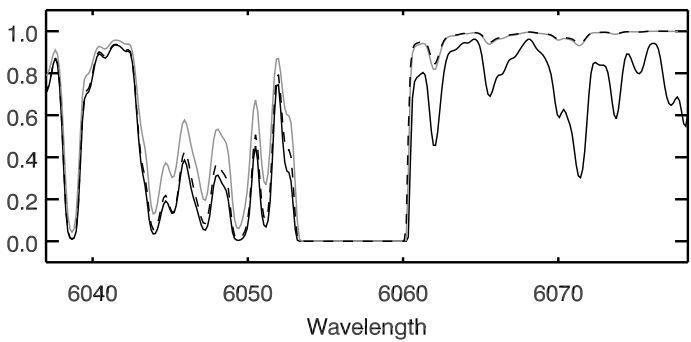}
\caption{\label{fig:sampSpecShad}Mock spectra synthesized from the $z=4$ simulation with
the QSO in a void. The solid line gives the Ly$\alpha$ forest
spectrum without the influence of the QSO, 
the dashed line the spectrum from the radiative transfer simulation
of the proximity effect for a QSO Lyman limit luminosity of 
$L_{\nu_{\rm LL}}=10^{31}\textrm{ erg
Hz}^{-1}\textrm{s}^{-1}$. The grey line gives results
from a semi-analytical model of the proximity effect showing strong deviations
from the radiative transfer results behind the strong absorber at the middle
of the spectrum. No noise is added to the spectra for better visibility.}
\end{figure}

We now focus on the over-ionisation field and discuss how optically thick regions in the intergalactic
medium affect the proximity effect. A slice of the over-ionisation field is shown 
in the right panel of Fig. \ref{fig:shadowingShielding} 
for the $L_{\nu_{\rm LL}}=10^{31}\textrm{ erg Hz}^{-1}\textrm{ s}^{-1}$ QSO
residing in a void at $z=4$. Again the smooth over-ionisation profile is seen as expected. 
However in this smooth transition zone, large shadow cones originating at dense regions
are present. The optical depth of these clouds is large enough to absorb 
most of the ionising photons produced by the QSO.
Hence the ionisation state of the medium behind such an absorber stays at the 
unaltered UVB level.

We determined the hydrogen column density of the shadowing regions
in Fig. \ref{fig:shadowingShielding}. They lie in the range of
$1\times10^{17} < N_{\ion{H}{i}} < 5\times10^{18} \textrm{ cm}^{-2}$. The regions causing the shadows
therefore represent Lyman limit systems. Whether such systems are able to absorb a large amount of QSO
photons depends on the one hand on their density and on the other on the QSO flux
reaching them. Absorbers further away from the source will receive less flux from the QSO 
due to geometric dilution. Therefore they are more likely to stay optically thick. For systems nearer to the QSO,
the flux field is larger and a higher density is needed for them to remain optically thick. Thus the number of shadows
increases with increasing distance to the source. Moreover a highly luminous QSO will produce
a larger amount of ionising photons and a higher number of absorbers are rendered optically thin. Hence
less shadows are present in the over-ionisation field of a highly luminous QSO.

By extracting a spectrum through such an absorber (see Fig. \ref{fig:sampSpecShad})
the influence of the Lyman limit system on the proximity effect signature can be seen. Behind
the strong absorber on the left hand side, the spectrum follows the unaffected Ly$\alpha$ forest spectrum. 
Measurements of the proximity effect behind such a system do not show any QSO influence.
This shadowing is not described with a semi-analytical model where the optical depth in the forest is
altered proportional to $1 + \omega(z)$. We will discuss and make use of this semi-analytical
model in Section
\ref{sec:LOSResult}. By looking at the semi-analytic model spectrum shown in Fig. \ref{fig:sampSpecShad} it is
clear that in this model the proximity effect extends beyond the Lyman limit system.

\subsection{\label{sec:Diffusion}Diffuse Recombination Radiation}

\begin{figure*}
\begin{minipage}[t][1\totalheight]{0.48\textwidth}
\includegraphics[bb=10bp 0bp 412bp 278bp,clip,
                 width=1\columnwidth]{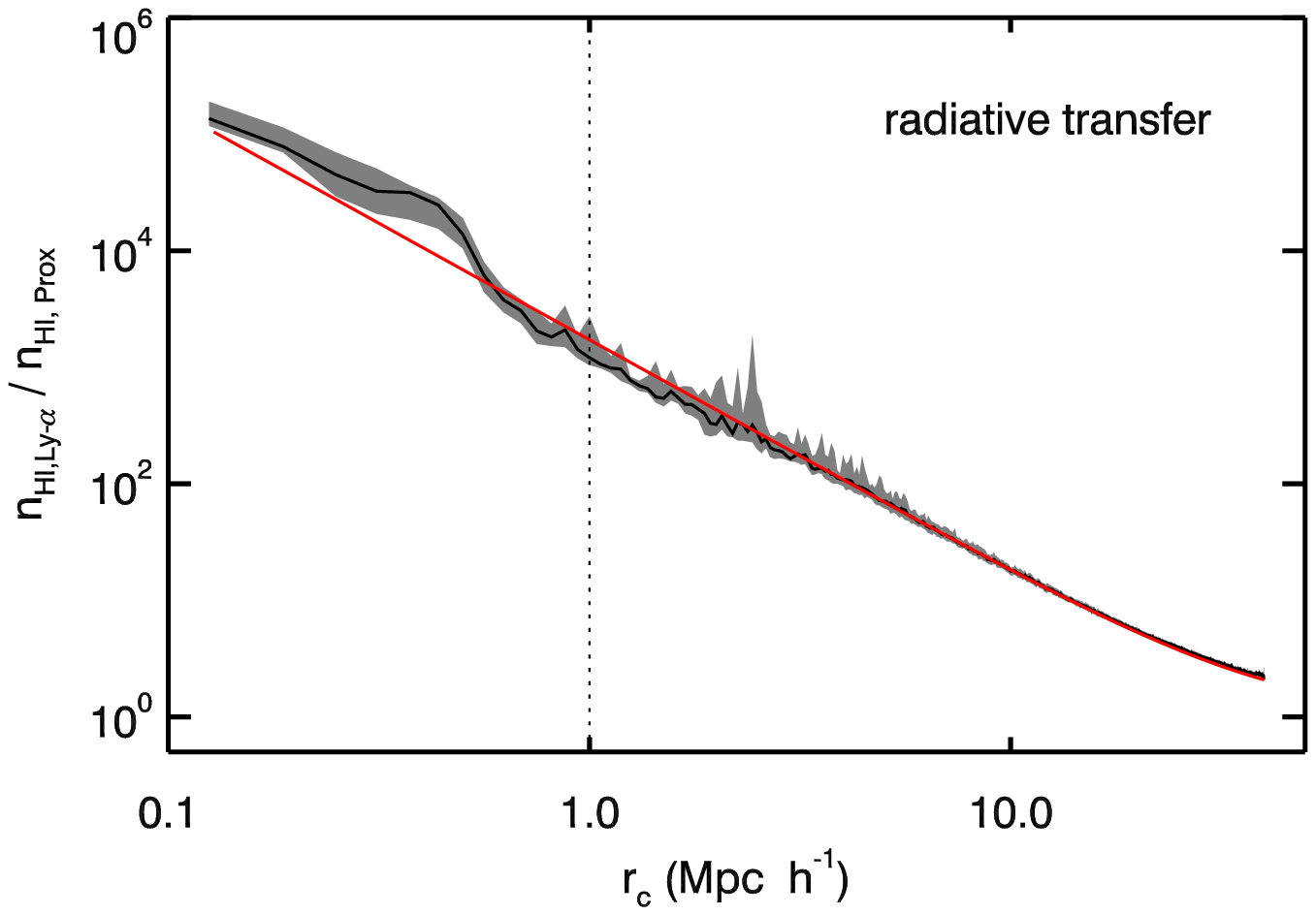}

\includegraphics[bb=10bp 0bp 412bp 278bp,clip,
                 width=1\columnwidth]{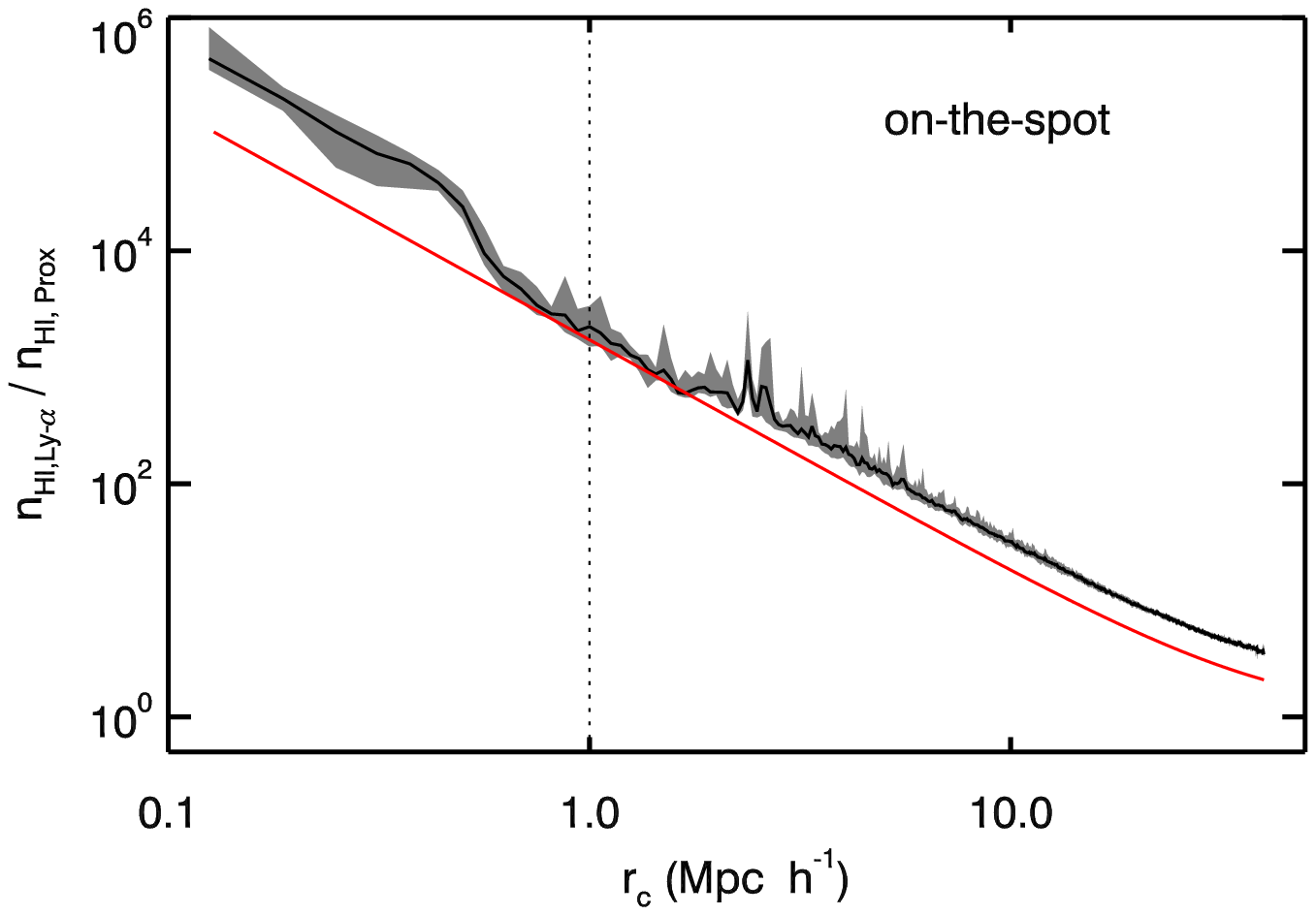}
\end{minipage}
\hfill{}
\begin{minipage}[t][1\totalheight]{0.48\textwidth}
\includegraphics[bb=10bp 0bp 412bp 278bp,clip,
                 width=1\columnwidth]{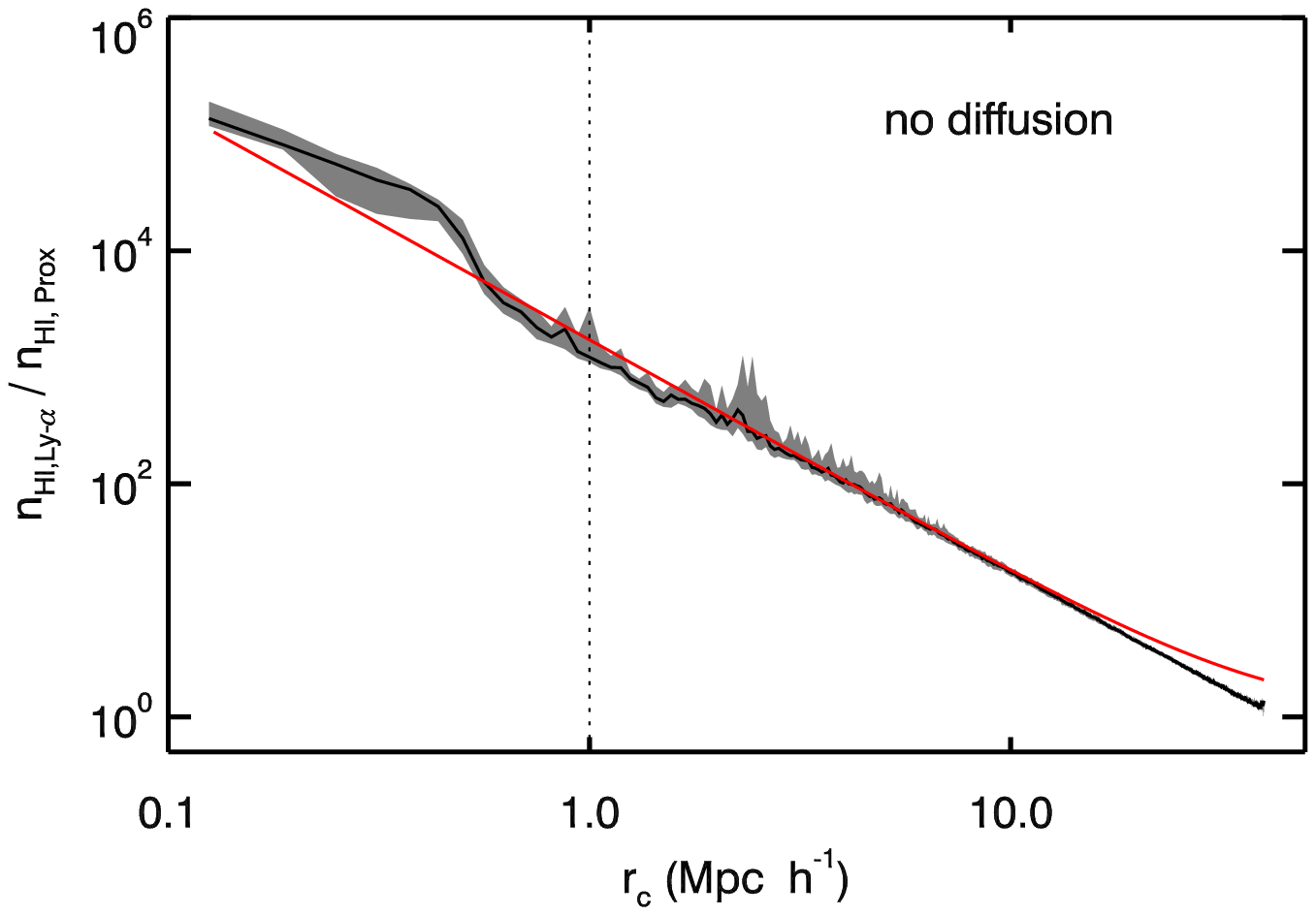}
                 
\includegraphics[width=0.99\columnwidth]{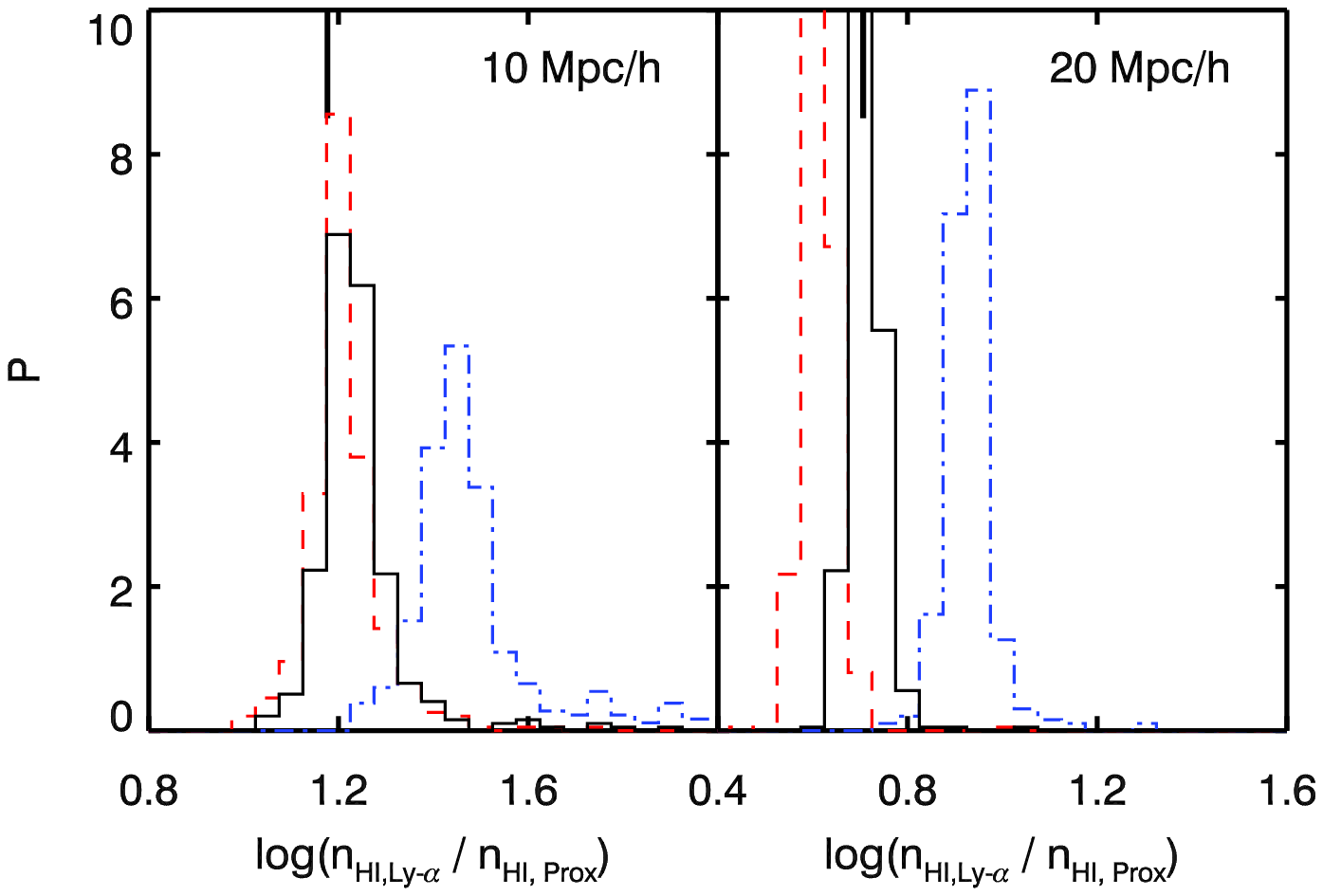}
\end{minipage}
\caption{\label{fig:scatterComp} 
Median over-ionisation profile as a function of the distance from the QSO 
hosted by a halo as in Fig. \ref{fig:ProxyProZ3}. We show
the influence of the diffuse field on the \ion{H}{ii} region for
$L_{\nu_{\rm LL}}=10^{32}\textrm{ erg Hz}^{-1}\textrm{ s}^{-1}$
at $z=3.0$. The upper left plot shows the full radiative transfer simulation, 
the right plot the one without diffusive radiation, and the lower left plot
results using the on-the-spot approximation. The analytical over-ionisation profile including the 
SED effects is given by the solid line. 
The region affected by oversampling ($r < 1 \textrm{ Mpc}$ ) is 
marked with a dotted line.
The lower right 
histograms gives the distribution of the over-ionisation fraction at
radii $r=10\;\textrm{ Mpc }h^{-1}$ (left) and $r=20\;\textrm{ Mpc }h^{-1}$ (right).
The solid lines give radiation transfer results, the dashed lines shows results 
omitting diffusion, and the dot-dashed distribution illustrates results obtained with
the on-the-spot approximation. The solid tick gives the analytical solution.}
\end{figure*}

We will now discuss the role played by photons produced by recombining
electrons with the help of the over-ionisation profile, since our
radiative transfer simulations give the opportunity to study this diffuse component. With
the Monte-Carlo scheme we can directly model the radiation field produced by recombination, which in 
most other schemes is not treated self consistently.
Diffuse radiation produced by recombining electrons plays an important role in radiative transfer problems
if the medium is optically thin. Whenever photons originating in recombination events are able to travel over large 
distances (i.e. larger than
one cell size), energy is redistributed on the scale of the photon's mean free path and the ionisation state is altered
over these distances.
The importance of diffuse radiation in the simplest of radiative transfer problems, the Str\"omgren sphere,
has been shown in
\citet{Ritzerveld:2005} where the outer 30\% of the Str\"omgren radius are found to be dominated by recombination
photons. Further \citet{Aubert:2008fk} used their radiative transfer code to discuss how much the on-the-spot 
approximation, a popular assumption that treats recombination photons as a local process, 
alters results in widely used test cases.

Recombination photons to the hydrogen ground level possess enough energy
to ionise other hydrogen atoms. If such a photon is absorbed by a nearby neutral
atom, the on-the-spot approximation can be applied where 
it is assumed that all recombination photons are
absorbed by neighbouring atoms. This is true if the medium is dense 
and the optical depth high enough that a local equilibrium between recombination 
and absorption is reached. If however the optical
depth is small, the mean free path is larger 
than the simulation cell size and the recombination photons are absorbed far away from 
their origin. In this case, the on-the-spot approximation breaks down.

We expect recombination radiation to play an important role in the proximity effect, since the optical depths
in the Ly$\alpha$ forest are low. To identify the region where recombination 
processes influence the solution, 
we run one simulation without emitting recombination photons.
However we still use case A recombination rates and call this the `no
diffusion case'. Furthermore we implemented the on-the-spot approximation 
using case B recombination rates. As before we
do not follow any recombination photons.
We expect to see differences at lower redshift, where the mean free paths for 
ionising photons
are large. Therefore we consider the snapshot at redshift $z=3$ with the
$L_{\nu_{\rm LL}}=10^{32}\textrm{ erg Hz}^{-1}\textrm{s}^{-1}$ QSO 
residing in a halo for this experiment. 

The resulting median over-ionisation profiles are presented in Fig. \ref{fig:scatterComp}. 
We also show the over-ionisation distribution at two radii to better compare the different 
runs. At radii $r > 1 \textrm{ Mpc } h^{-1}$ we obtain profiles that well reproduce the
theoretical estimates. We see some noisy fluctuations in the profiles.
Due to the larger ionisation fractions for the bright QSO, the ionisation fractions themselves
are more susceptible to Monte-Carlo noise. Since the neutral hydrogen fractions in 
the direct vicinity of the QSO
are 2 dex lower for the bright QSO than for the faint one, higher precision is needed to 
better evaluate 
the extremely low neutral fractions. However further increasing the number of photon 
packages is at the moment beyond the capabilities of our code.

Comparing the no diffusion run with the full radiative transfer solution shows
that recombination greatly contributes to the outer parts of the proximity
effect region. The no diffusion solution does not gradually go over to unity, but keeps on
decaying fast until the QSO's influence vanishes. At $r=20\;\textrm{ Mpc }h^{-1}$ the
difference between the result with and without the diffuse recombination field
is 23\%. This means, that recombination radiation contributes to the photoionisation
rate by 23\% at this radius, if one assumes that the ionisation fraction is directly proportional
to the photoionisation rate.

The on-the-spot solution on the other hand over predicts the extent of the proximity effect zone
by about a factor of 2. This means that keeping the energy produced by 
recombining electrons locally confined over predicts the amount of ionised hydrogen. 
As a result, the QSO's sphere of influence is overestimated. Further the on-the-spot solution 
shows larger dispersion between 
the lines of sight than the radiative transfer simulation (see Fig. \ref{fig:scatterComp} lower 
right panel). We therefore conclude that the on-the-spot 
approximation is not sufficient when modelling the proximity effect.

\section{\label{sec:LOSResult}Radiative Transfer on the Line-of-Sight 
Proximity Effect}

\subsection{\label{sec:StrengthParam} Strength Parameter}

 In this Section we discuss the imprint of radiative transfer on the
line-of-sight proximity effect in Ly$\alpha$ spectra as analysed in observations.
To this aim we study a sample of 500 lines of sight originating at the QSO position and
having randomly drawn directions. We then measure the proximity effect signature.

First an appropriate $\omega$-scale is constructed
for each QSO applying Eq. \ref{eq:omega2} including the SED correction term. In Section
\ref{sec:SEDPerf} we discuss how well the correction term accounts for the
SED effect. Given the $\omega$-scale we then determine the mean transmission
in bins of $\Delta \log \omega = 0.5$ for each line of sight. The mean proximity effect signal 
is calculated by averaging the mean transmission in each $\log \omega$ bin
over the QSO lines of sight. We determine the mean optical depth in each bin. 
The relative change introduced in the optical depth by the QSO is evaluated by 
computing the normalised optical depth $\xi$ defined as
\begin{equation}
\xi=\frac{\tau_{{\rm eff,\, QSO}}}{\tau_{{\rm eff,\, Ly\alpha}}}, 
\label{eq:xi}\end{equation}
where $\tau_{{\rm eff,\, QSO}}$ is the effective optical depth in a $\log \omega$ bin and
$\tau_{{\rm eff,\, Ly\alpha}}$ the effective optical depth in the Ly$\alpha$ forest 
without the influence of the QSO.
For $\Delta \log \omega < 1$ we find the resulting normalised optical depths to be 
weakly dependent on the chosen bin size.

To quantify any difference in the proximity effect signature to the expected one given by
Eq. \ref{eq:omega2}, we adopt the `strength parameter' as in
\citet{DallAglio:2008uq, DallAglio:2008ua}. 
The strength parameter $a$ is defined as
\begin{equation}
\xi=\left(1+\frac{\omega}{a}\right)^{1-\beta} \label{eq:omegaA}
\end{equation}
where $\omega$ is defined by Eq. \ref{eq:omega2}.
 As we know the UVB
photoionisation rates in our models, we use the values given in Table \ref{tab:gnedModels}
to define the reference $\omega$. Then values $a > 1$ or $a < 1$  
describe weaker or stronger proximity effects, respectively. 
The SED correction term in Eq. \ref{eq:omega2} 
can also be interpreted in the framework of the proximity strength parameter.

It is also possible to determine the strength parameter for each line of sight individually.
By obtaining the proximity effect strength for each line of sight, the strength parameter 
distribution function can be constructed. Studying its properties gives further insight 
into the proximity effect signature and has been used by \citet{DallAglio:2008uq} to 
derive the UVB photoionisation rate. How well this method
performs in the context of the present study will be discussed in Section \ref{sec:PESD}.

\subsection{\label{sec:PEModels}Additional Models}

 In addition to the lines of sight obtained from the radiative transfer simulation, we explored
two other models allowing us to characterise the imprint of the cosmological
density distribution and radiative transfer in the analysis of the proximity effect.
These complementary analyses will be used to obtain spectra that can be directly 
compared to our radiative transfer results.

\paragraph{\label{sec:SAM}Semi-Analytical Model (SAM) of the Proximity Effect:}

From the radiative transfer simulation of only the UV background, we know the
ionisation state of hydrogen in the box without the influence of the QSO. 
Then we can semi-analytically 
introduce the proximity effect along selected lines of sight from the QSO 
by decreasing the neutral fractions by the factor 
$(1+\omega)$. We choose the same 500 lines of sight as in the full 
radiative transfer analysis.

This model includes the imprint of the fluctuating density field on the spectra.
Since the semi-analytical model only includes geometric dilution and the
influence of the SED, 
any differences to the results of the full radiative transfer
simulations reveals additional radiative transfer effects.

\paragraph{\label{sec:RAM}Random Absorber Model (RAM):}  

In our radiative transfer simulations, we can only cover the proximity effect signature up to 
$\log \omega \ge 0.5$ for the faint QSO and to $\log \omega \ge 1.0$ for the strong one, due
to the finite size of our simulation box. In order to overcome this limitation, we employ a 
simple Monte Carlo method 
to generate Ly$\alpha$ mock spectra used in observations to study systematic effects 
in the data \citep{Fechner:2004kk, Worseck:2007wq}.
The Ly$\alpha$ forest is randomly populated with  
absorption lines following observationally derived statistical properties. The constraints
used are: 1) The line number density distribution, approximated by a power  
law of the form $\mathrm{d}n / \mathrm{d}z \propto (1+z)^{\gamma}$  
where $\gamma=2.65$ \citep{Kim:2007kx}, 2) the column density distribution, given by 
$f(N_{\ion{H}{i}})  \propto N_{\ion{H}{i}}^{-\beta}$ where the slope is $\beta \simeq  
1.5$ \citep{Kim:2001fk}, and 3) the Doppler parameter distribution, given by  
$\mathrm{d}n /  \mathrm{d}b \propto b^{-5} \; \mathrm{exp}\left[{-{b_{\sigma}^4}/{b^4}} 
\right]$, where  $b_{\sigma}\simeq 24\;\mathrm{km/s}$ \citep{Kim:2001fk}.
For a detailed discussion of the method used, see \citet{DallAglio:2008ua, DallAglio:2008uq}.

The random absorber model does not take clustering of absorption lines due 
to large scale fluctuations in the density distribution into account . Therefore we use this model
to determine any possible bias in the analysis method and to infer the effect of absorber clustering. 
Furthermore, since we can generate 
spectra with large wavelength coverage as in observated spectra, we can quantify the effect 
introduced by the truncated coverage of the $\omega$ scale in the spectra from the 
radiative transfer simulation. With this simple model it is possible to study any 
systematics present in the analysis, such
as the influence of the $\Delta \log \omega$ bin size on the resulting proximity effect.

\section{\label{sec:meanProx}Mean Proximity Effect}

Now we discuss the results of the mean proximity effect as measured in the synthesized
spectra.

\subsection{\label{sec:moreMeanProx}Mean Normalised Optical Depth}

The main results are presented in Fig. \ref{fig:ProxyMultiOmega15} and 
Fig. \ref{fig:ProxyMultiOmega}. In both plots we show the analytical model according to 
Eq. \ref{eq:tauProfile} and Eq. \ref{eq:xi} for comparison. The model predicts that at 
high $\omega$ values, the QSO shows the strongest influence and the normalised optical 
depth is very low (at the QSO where $\omega = \infty$, the normalised optical depth 
is $\xi = 0$). Then between $2 > \log \omega > -1$ the influence of the QSO
gradually declines due to geometric dilution and reaches $\xi = 1$ for small $\omega$ values.
There the Ly$\alpha$ forest is unaffected by the QSO.

The normalised optical depth profile derived from the transfer simulations closely
resembles the functional form of the analytical expectations.
However a shift of the profile is present which we quantify using the proximity strength
parameter. In observations, the strength parameter is used to measure the UVB
photoionisation rate. We therefore faithfully apply the same method to the simulation results.
However, since we know the UVB in our models, we can directly compare the photoionisation
rate determined using the strength parameter with the model input.
The strength parameter is determined by fitting Eq. \ref{eq:omegaA} to the data points 
using the Levenberg-Marquardt algorithm and a least absolute 
deviation estimator weighted with the error of the mean. In order to obtain a reasonable 
fit, data points deviating strongly from 
the profile above $\log \omega > 2$ have been excluded. 
This also excludes the unreliable region within 1 Mpc of the QSO in the radiation transfer simulation, 
which translates to an $\log \omega \ge 2.4$ for the faint QSO and $\log \omega \ge 3.4$ for the
bright one.
We provide the resulting strength parameters in Table \ref{tab:aParamTable} and include the fitted 
profiles in all plots of the mean proximity effect. The origin of the shift is due to the
large scale environment around the QSO hosts which we discuss in more details
in Section \ref{sec:largeScaleEnv}.

For each data point we show two error bars, the standard deviation of the sample and the 
$2\sigma$ error of the mean. The sample standard deviation gives the variance between the 500
lines of sight. These variances are 
very large and extend to values above $\xi > 1$. Values of $\xi > 1$ arise when
saturated absorption systems dominate the optical depth in a $\log \omega$ bin and 
raise the optical depth over the mean effective optical depth in the forest. 
A detailed illustration is given in the Appendix. Since the real space distance 
covered by the $\log \omega$ bins increases with decreasing $\omega$, there the contribution 
of saturated systems and the variance is reduced. 

The error of the mean is on the other hand small due to our sample size of 500 lines of sight, 
thus the mean profiles are well determined. 
However the mean profile does not strictly follow the smooth analytical model, but shows
strong fluctuations. 
These fluctuations are enhanced by the fact that we can only analyse the QSO in one octant. Further
we are focusing our analysis just on a single object, making the signal sensitive to the surrounding
density distribution. If multiple halos would be combined to one mean profile, the fluctuations would
diminish and the profile would follow the smooth analytical profile more closely.
Since the data points deviate from the analytical profile, the determination of the strength 
parameter is quite uncertain.
To estimate a formal error in the strength parameter we applied the Jackknife method. 
The error values are provided in Table \ref{tab:aParamTable}. 

We checked if Monte-Carlo noise of the radiation transport simulation contributes to the signal's 
variance by reducing the QSO's photon packages production by half. This does not
contribute noticeably to the variance in the signal.
The large scatter in the proximity effect signal therefore has its origins
in the distribution of absorption systems along the line of sight. It is 
a direct imprint of the cosmic density inhomogeneities.
The probability distribution of the normalised optical depths in each 
$\log \omega$ bin is approximately log-normal as shown in Fig. \ref{fig:xiDistribution}. 

\subsection{\label{sec:SEDPerf} Performance of the SED Correction Term}

\begin{figure}
\centering
\includegraphics[width=1.0\columnwidth]{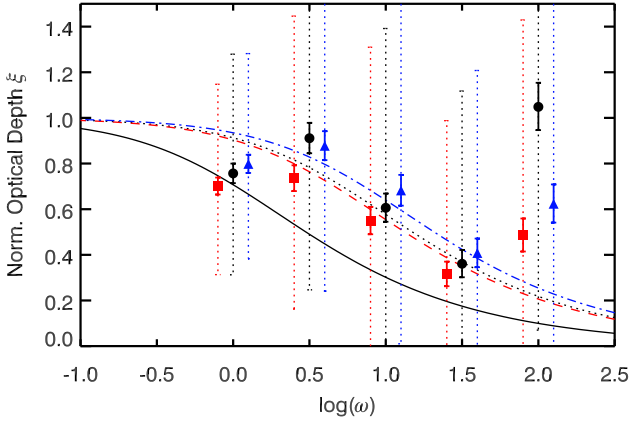}

\caption{\label{fig:ProxyMultiOmega15}The normalised optical depth for the 
$L_{\nu_{\rm LL}}=10^{31}\textrm{ erg Hz}^{-1}\textrm{ s}^{-1}$ QSO with the standard SED 
and residing in a halo at
$z=3$. Continuous lines show the fitted profiles. The triangles and dash-dotted line 
give the radiation transfer proximity effect without the use of the SED correction term. 
The circles and dotted line give the radiation transfer results including the SED correction. 
The QSO with the hard $\alpha_\mathrm{q}=-1.5$ SED is given by the squares and 
dashed line. The original BDO model
is shown as the solid line. The dotted error bars show the sample standard deviation.
The solid errors show the $2\sigma$ error of the mean.}
\end{figure}

With our radiative transfer simulations it is possible to see, whether the SED 
correction term in Eq. \ref{eq:omega2} describes the effect of a QSO and UV background
which have different spectral shapes. The SED 
adds to the mentioned shift in the normalised optical depth, $\xi$, already
introduced by the large scale density distribution around 
the QSO hosts. In order to single out the SED influence, we re-simulate the $z=3$ 
faint QSO residing in a halo with a $\alpha_\mathrm{q}=-1.5$ power-law SED. 
With the background's SED slope of $\alpha_\mathrm{b} = -1.28$, we expect only
a slight shift due to the different slopes of $\Delta \log a = 0.02$. The softer 
\citet{2007AJ....133.1780T} SED used in the rest of the paper yields a shift of 
$\Delta \log a = 0.11$. If the SED correction term performs as intended, the corrected
proximity effect profile of the soft SED QSO would approximately lie on top of the one with 
the hard SED.

In Fig. \ref{fig:ProxyMultiOmega15} we present the results of this experiment, where we plot
the proximity effect profile of the $\alpha_\mathrm{q}=-1.5$ QSO, and SED corrected and 
uncorrected profile for the QSO with $\alpha_\mathrm{q} \approx -2.5$. The data points of the
corrected $\alpha=-2.5$ profile lie almost on the $\alpha_\mathrm{q}=-1.5$ profile as expected. 
A similar picture emerges by looking at the strength parameters.  
The uncorrected soft SED yields a strength parameter of $\log a = 0.84 \pm
0.06$.  
By including the correction term we obtain $\log a = 0.70 \pm 0.12$.
For the $\alpha_\mathrm{q}=-1.5$ SED we obtain a strength parameter of 
$\log a = 0.65 \pm 0.05$. The different values of the SED corrected and the soft QSO 
results are consistent within the error bars. 

\subsection{\label{sec:largeScaleEnv} Influence of the Large Scale Environment}

\begin{figure}
\centering
\includegraphics[width=1.0\columnwidth]{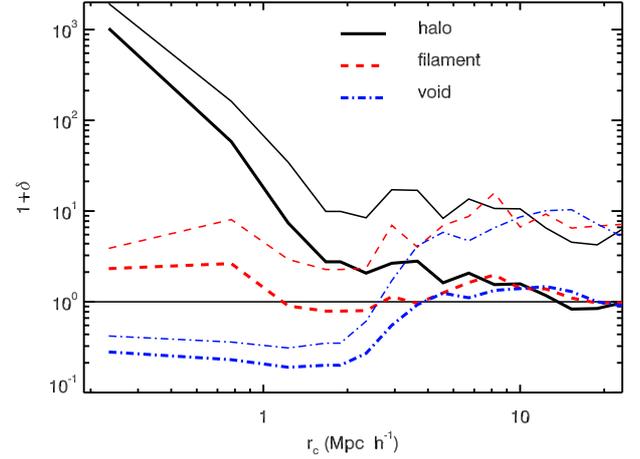}

\caption{\label{fig:densityEnv} The mean overdensity profiles of the halo
  (solid line), filament (dashed line), 
and void (dash-dotted line) environment at redshift $z=3$ as a function of comoving radius. 
The mean profiles are shown by thick lines. 
The thin lines give the $1\sigma$ fluctuations in the density field.
An excess in overdensity is seen for all environments between $3 < r_c < 10 \textrm{ Mpc } h^{-1}$, where
the profile stays above the average overdensity.}
\end{figure}

The environmental density around the QSO has been shown to strongly influence
the proximity effect. By studying random DM halos in a mass range of  
$1.4 \times 10^{11} M_{\sun} h^{-1}$ 
up to $9 \times 10^{12} M_{\sun} h^{-1}$, \citet{Faucher-Giguere:2008gf} have found 
an enhancement in the mean overdensity profiles up to a comoving radius of 
$(10 - 15) \textrm{ Mpc } h^{-1}$. Only at larger radii the average
overdensity profiles goes over to the
cosmic mean. The overdensity profiles fluctuate strongly between
different halos. Since an enhanced density directly translates into an enhancement
of the Ly$\alpha$ forest optical depths, the proximity effect 
measurement is biased accordingly.

\begin{figure*}
\begin{minipage}[t][1\totalheight]{0.49\textwidth}%
\includegraphics[bb=10bp 0bp 412bp 278bp,clip,width=1\columnwidth]
                {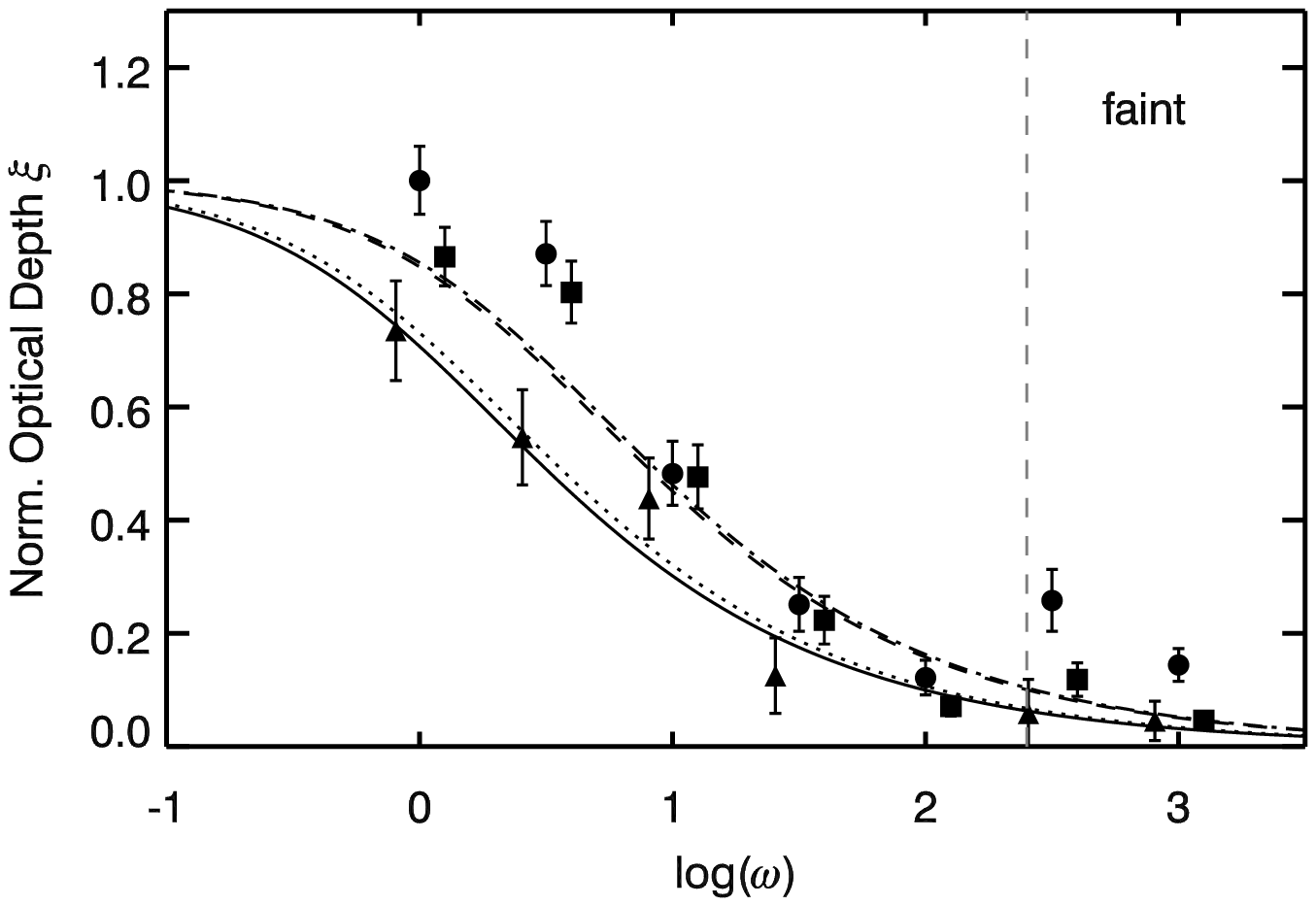}%
\end{minipage}%
\hfill{}%
\begin{minipage}[t][1\totalheight]{0.49\textwidth}%
\includegraphics[bb=10bp 0bp 412bp 278bp,clip,width=1\columnwidth]
                 {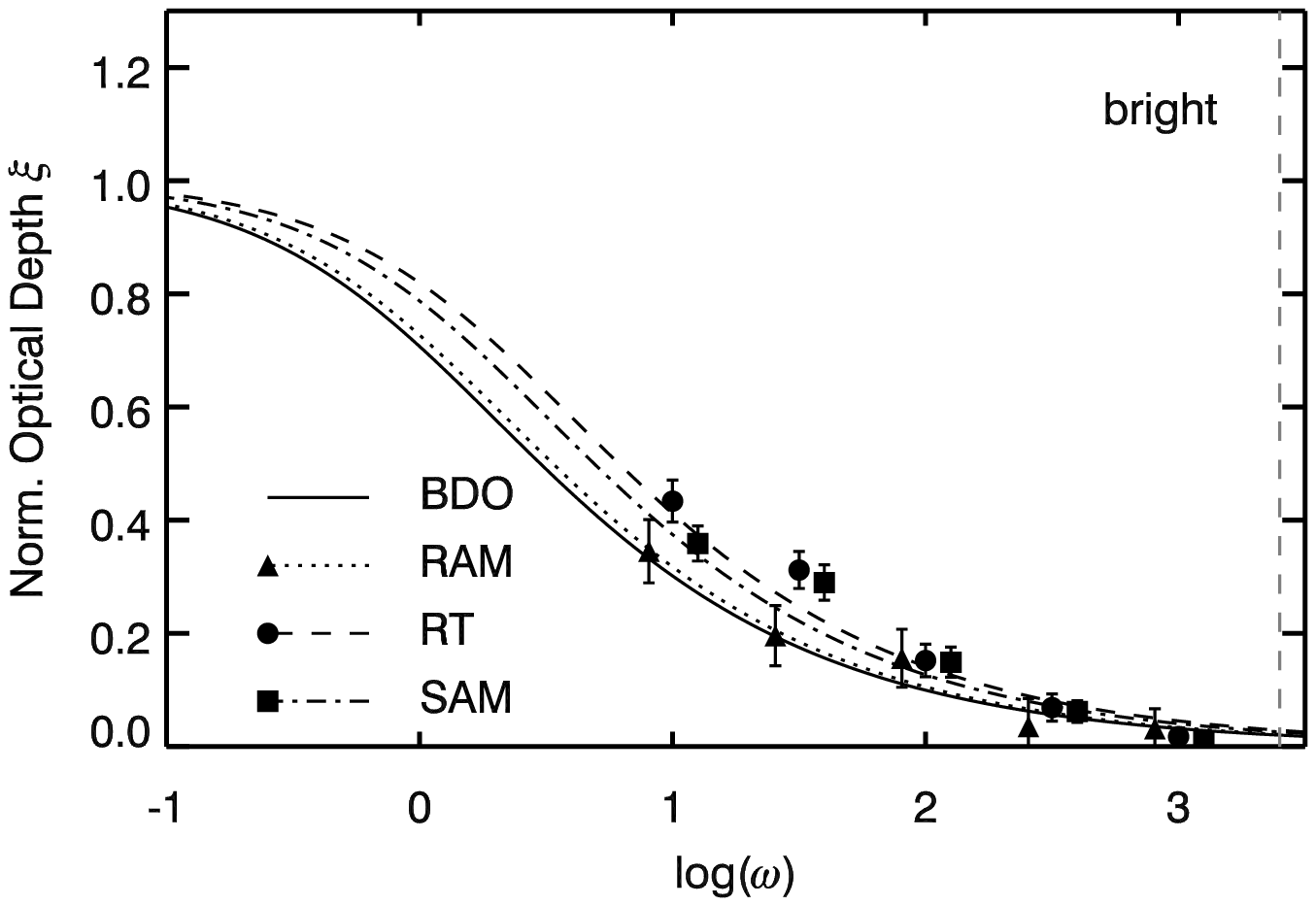}%
\end{minipage}%
\caption{\label{fig:ProxyMCOmega}Comparison of the proximity effect  
at redshift $z=3$ of a) the RAM (triangles and dotted line), 
b) the semi-analytical model (circles and dash dotted line), and
c) the results from the radiative transfer simulation (squares and dashed
line). The solid line shows the BDO reference analytical model. The dashed
vertical line marks the start of the unreliability region of $\log \omega \ge 2.4$ for the faint QSO
and $\log \omega \ge 3.4$ for the bright QSO. 
Results are shown for a QSO sitting in a filament.  
The left panel is for the low luminosity source, and the right one for 
the high luminous one. We only show the $2\sigma$ errors of the mean.}
\end{figure*}

We have determined the mean overdensity profiles together with the
fluctuations around the mean for our three QSO host environments. The resulting mean
profiles are shown in Fig. \ref{fig:densityEnv} for $z=3$. Our density profiles show two
different influences. 

First, the local environment around the QSO host leaves an imprint on the profiles up to a 
radius of $r \approx 2 \textrm{ Mpc } h^{-1}$. This is seen in the halo case
where the massive host halo is responsible for a strong overdensity which steadily declines 
up to $r \approx 2 \textrm{ Mpc } h^{-1}$. The contrary is seen in the void case, 
where up to $r \approx 3 \textrm{ Mpc } h^{-1}$ the underdense region of the void is 
seen. At larger radii, the density increases quite strongly.

All our environments show an overdensity between
$r \approx 3 \textrm{ Mpc } h^{-1}$ up to a radius of around 
$r \approx 15 \textrm{ Mpc } h^{-1}$. In this region the density lies above the cosmic mean. 
We refer to this phenomenon as a large scale overdensity. Further away from the QSO, 
the cosmic mean density is reached. The same behaviour is seen at higher redshifts. 
There, however, the amplitude of the large scale overdensity is lower.
The presence of the large scale overdensity in the case of the most massive halo is a direct
consequence of the fact that the most massive halo forms where there is a large overdensity.
However around the filament and the void such a large scale overdensity is a random 
selection effect.

An excess in density translates into an excess of Ly$\alpha$ optical depth in this region, when 
compared to the mean optical depth. Thus the normalised optical depth $\xi$ increases and
leads to a biased proximity effect. The immediate environment around
the quasar goes out to $\log \omega \approx 1.5$ for the faint source and to
$\log \omega \approx 2.5$ for the bright one. At smaller $\log \omega$
the large scale overdensity influences the proximity effect signal up to 
$\log \omega \approx 0.3$ for the faint and to $\log \omega \approx  1.3$ for the bright QSO.

Motivated by these considerations we study the influence of the cosmic density fluctuations on 
the proximity effect profile. To understand whether radiative transfer effects or the large scale 
overdensity are responsible for the shift in the normalised optical depth shown in Fig. 
\ref{fig:ProxyMultiOmega15}, 
we make use of the reference models introduced in Section \ref{sec:PEModels}. 

\begin{figure*}
\begin{minipage}[t][1\totalheight]{0.49\textwidth}%
\includegraphics[bb=10bp 0bp 412bp 278bp,clip,width=1\columnwidth]{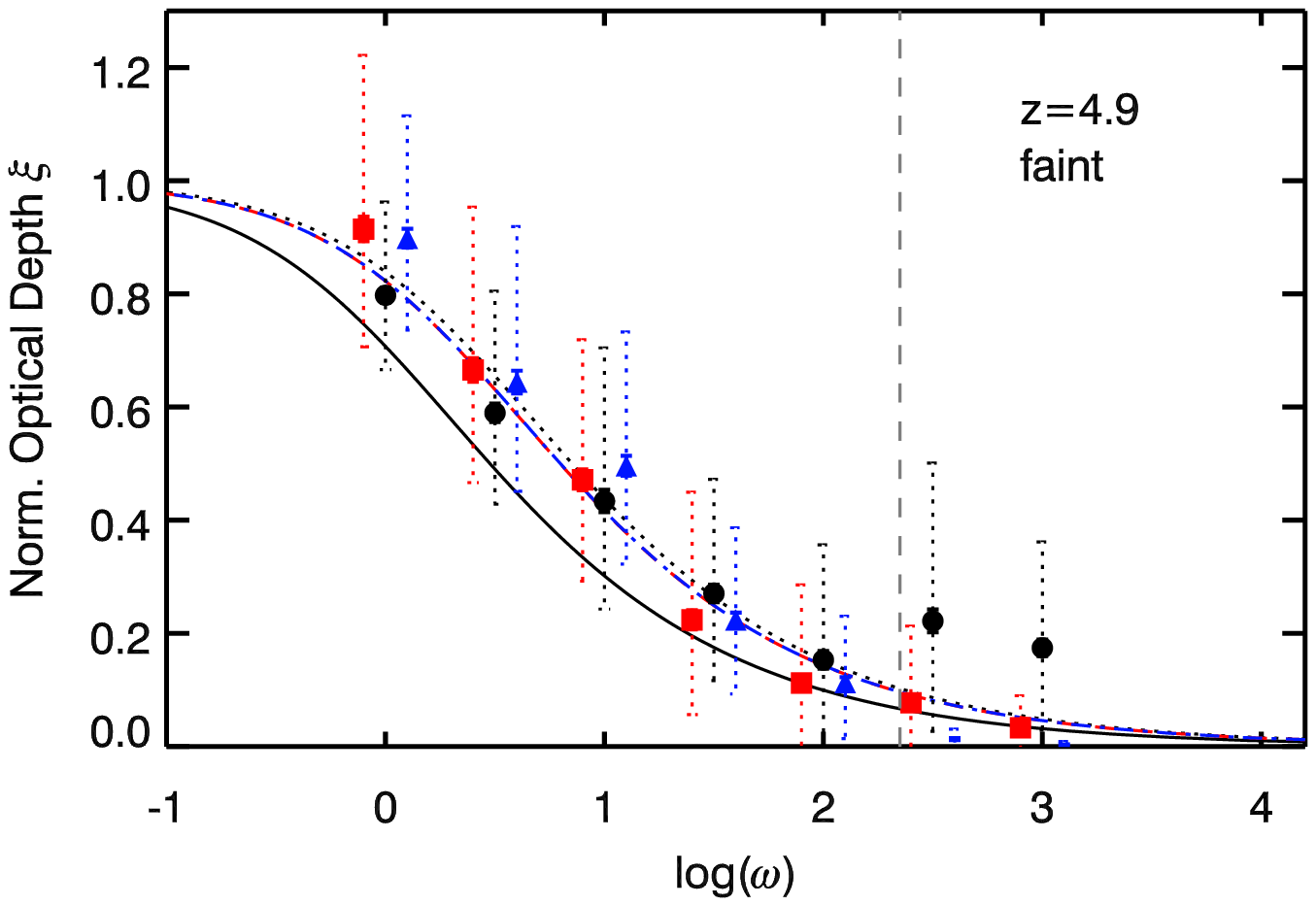}
\includegraphics[bb=10bp 0bp 412bp 278bp,clip,width=1\columnwidth]{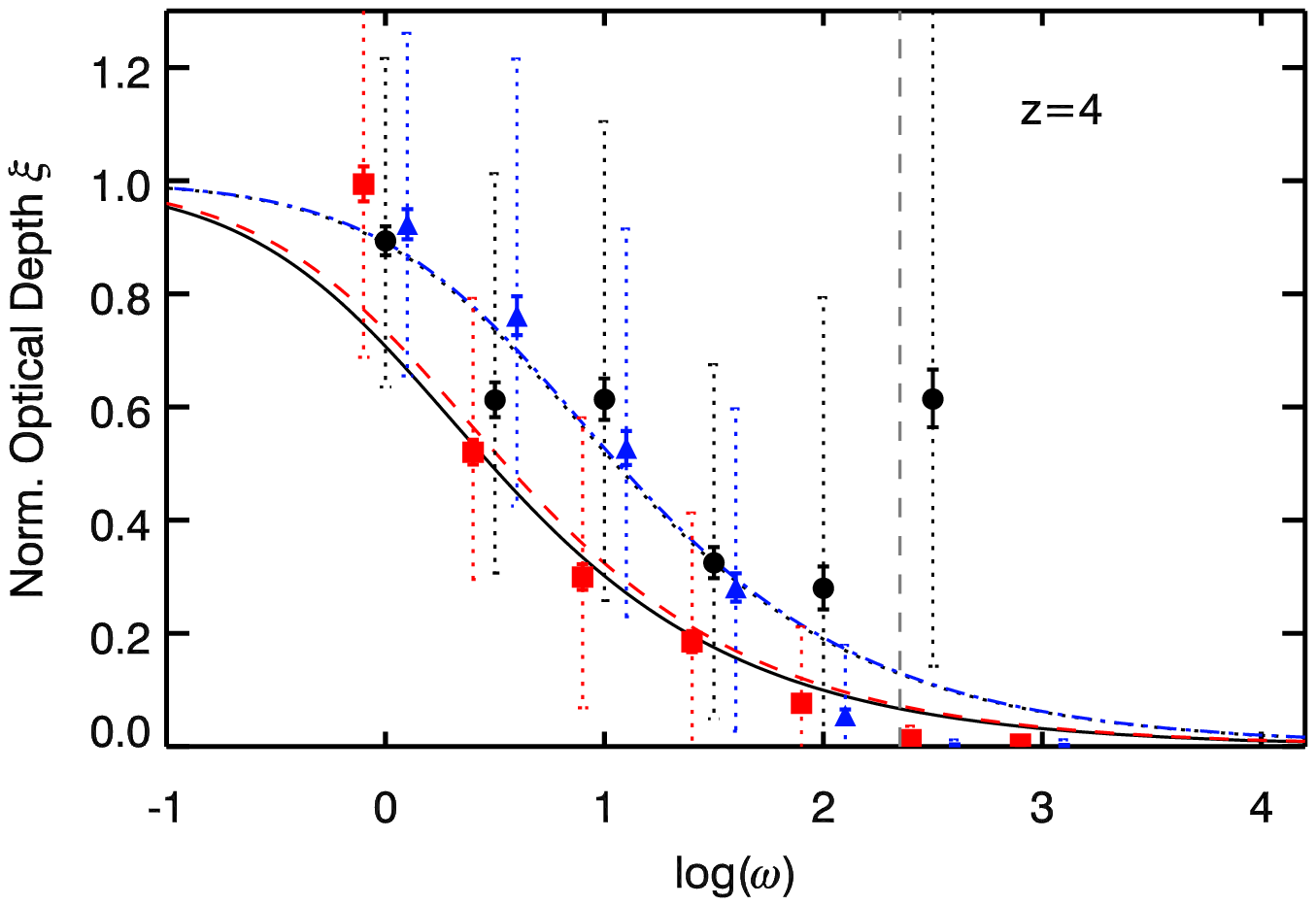}
\includegraphics[bb=10bp 0bp 412bp 278bp,clip,width=1\columnwidth]{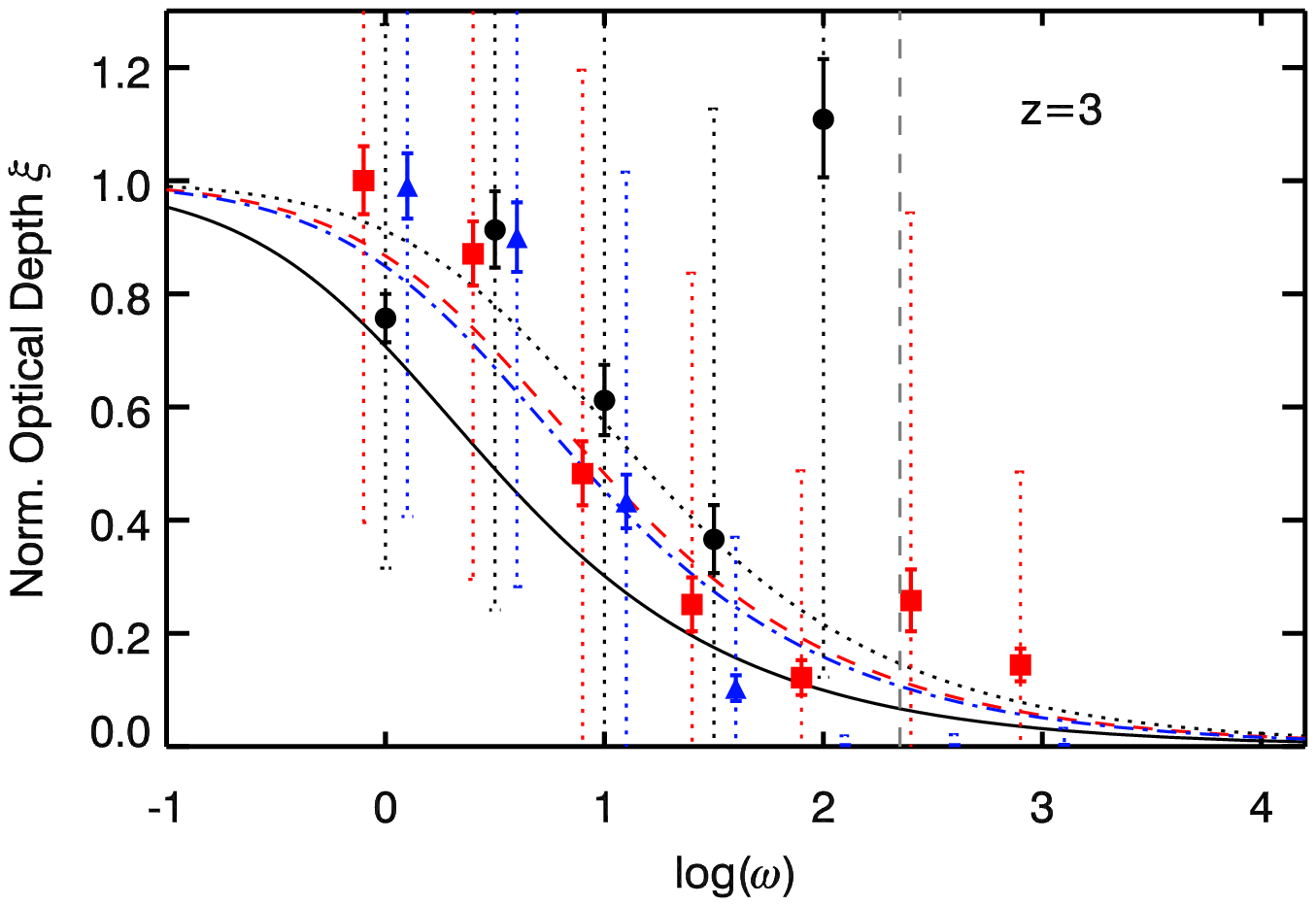}
\end{minipage}%
\hfill{}%
\begin{minipage}[t][1\totalheight]{0.49\textwidth}%
\includegraphics[bb=10bp 0bp 412bp 278bp,clip,width=1\columnwidth]{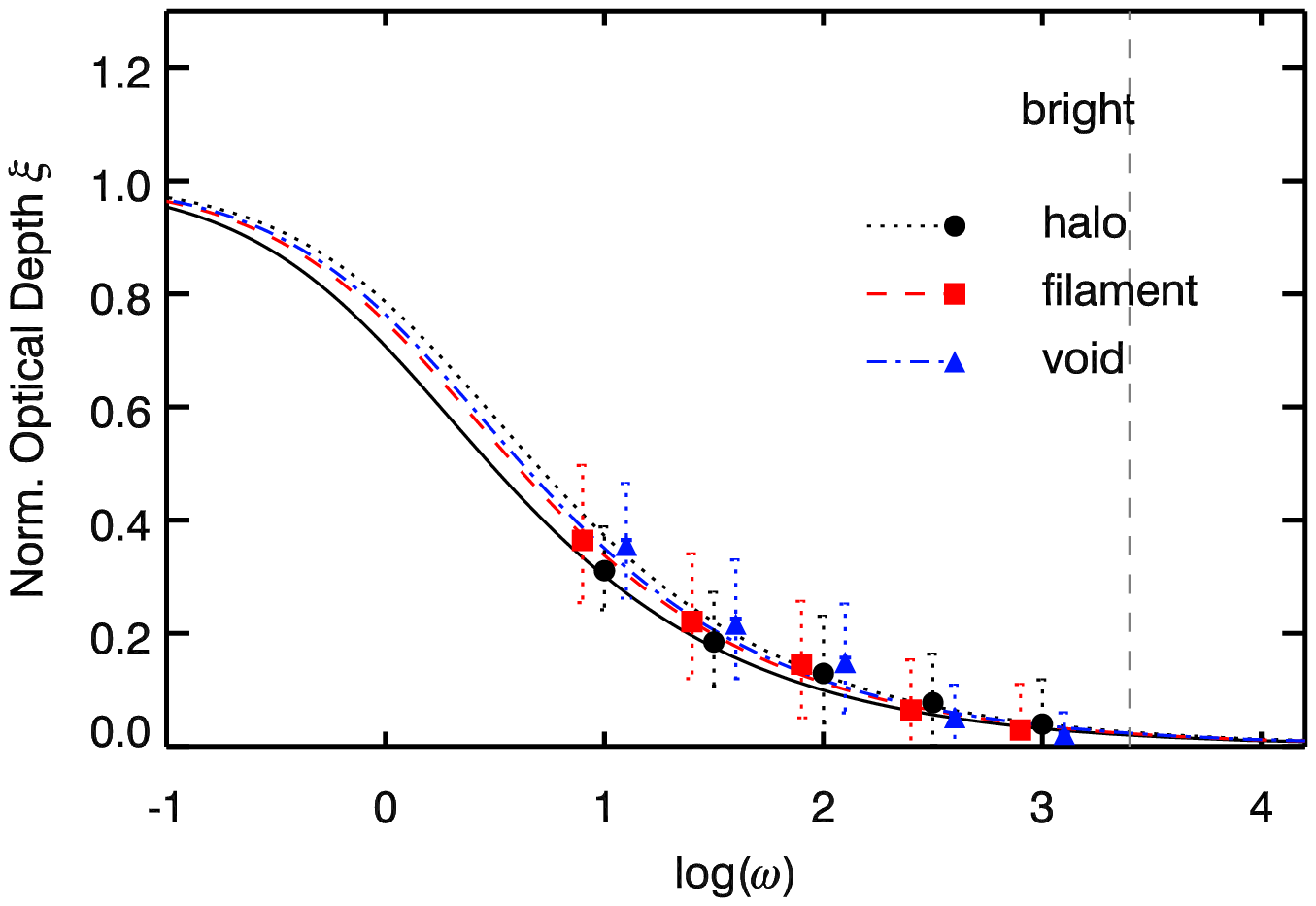}
\includegraphics[bb=10bp 0bp 412bp 278bp,clip,width=1\columnwidth]{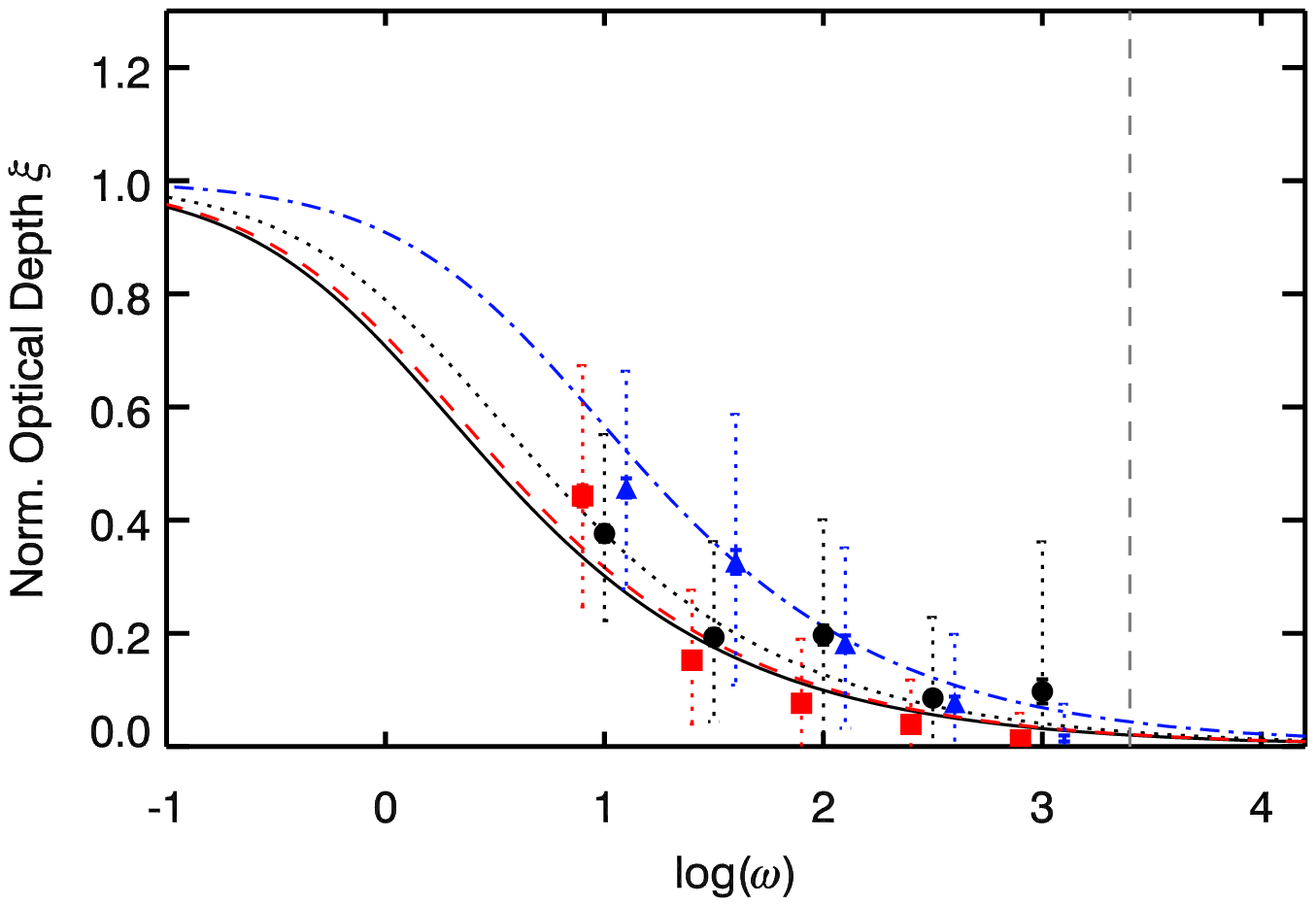}
\includegraphics[bb=10bp 0bp 412bp 278bp,clip,width=1\columnwidth]{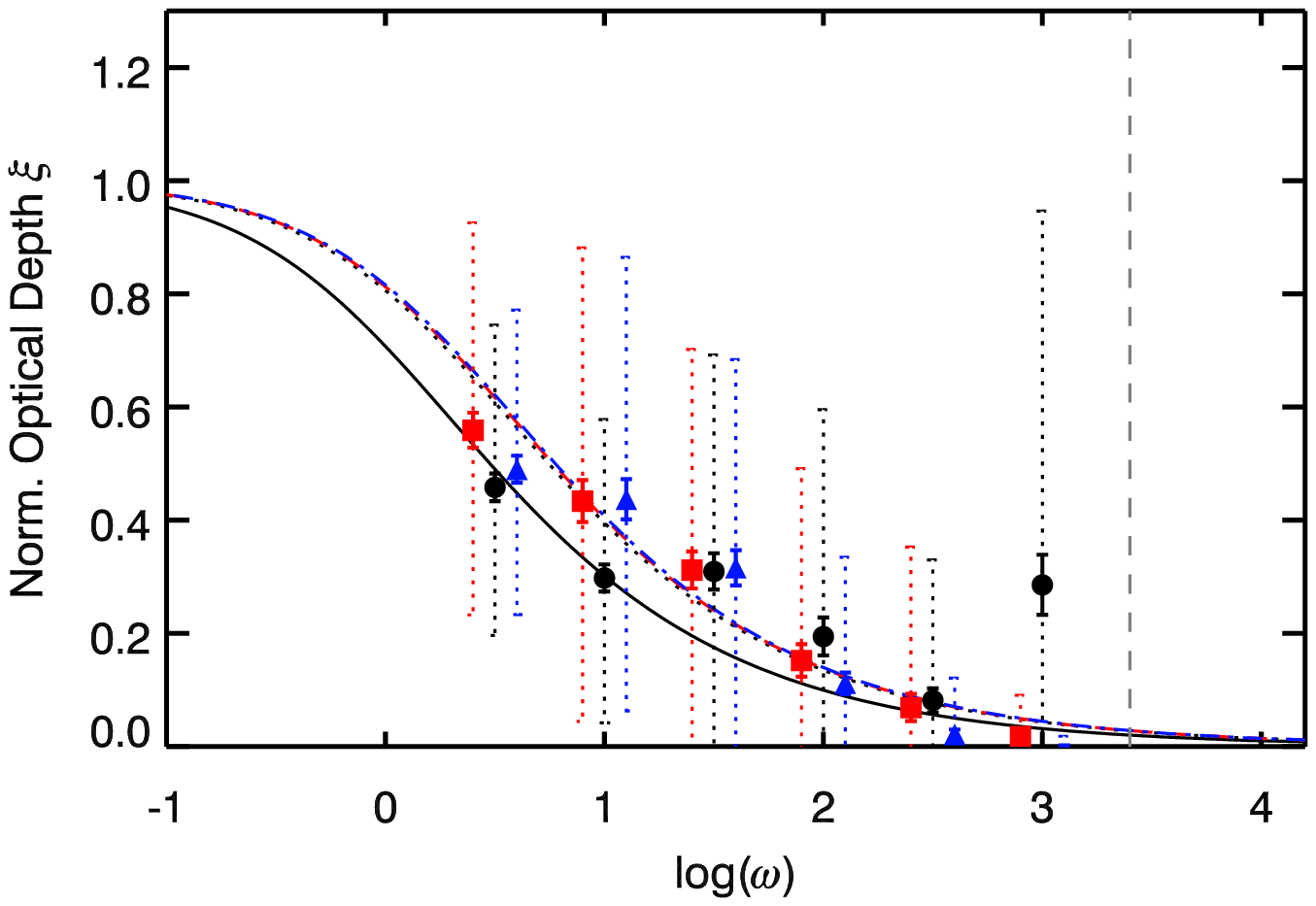}
\end{minipage}%

\caption{\label{fig:ProxyMultiOmega}The combined proximity effect
signature of all 500 lines of sight analysed with the BDO model showing the 
dependence on environment, luminosity and redshift. The left column shows
the case of a faint
$L_{\nu_{\rm LL}}=
10^{31}\textrm{ erg Hz}^{-1}\textrm{ s}^{-1}$ QSO, 
the right one of a luminous
$L_{\nu_{\rm LL}}=
10^{32}\textrm{ erg Hz}^{-1}\textrm{ s}^{-1}$ QSO.
From top to bottom redshifts are 4.9, 4, and 3. The solid 
lines show the geometrical dilution model, the other lines are best
fits to the measurements with parameters given in Table
\ref{tab:aParamTable}. The circles and dotted lines show the halo cases,
the squares and dashed lines the filament case and triangles and
dash-dotted lines the void case. Symbols are horizontally shifted by
$\pm0.05$ for better visibility. The dotted error bars show the sample $1\sigma$ 
standard deviation. 
The solid errors give the $2\sigma$ error of the mean.
The dashed vertical line marks the start of the unreliability region of 
$\log \omega \ge 2.4$ for the faint QSO and $\log \omega \ge 3.4$ for the bright QSO.}

\end{figure*}

We show the resulting proximity effect profiles for the reference models of the faint and bright QSO
residing in the filament at $z=3$ in Fig. \ref{fig:ProxyMCOmega}. We use the
RAM to infer the bias introduced by the analysis method itself due to the finite
size of our spectra. Additionally the SAM allows us 
to determine the influence of the fluctuating density field by omitting any possible radiative 
transfer effect. On top of that we show the radiative transfer results. 

The data points of the RAM follow the standard BDO profile well for both luminosities. However
a slight shift is seen in the fitted profiles. The strength parameters are 
$\log a = 0.06 \pm 0.05$ in the case of the weak QSO and $\log a = 0.05 \pm 0.10$ for 
the strong QSO. Within the error bars, this is consistent with the analytical predictions. 
A small
shift in the signal has been found to be intrinsic to the analysis method due to the asymmetry 
in the $\xi$-distribution \citep{DallAglio:2009lr}.

As for the radiative transfer results, the SAM shows a strong shift towards the QSO. 
This is a clear signal, that the cosmological density distribution is responsible for the large shift.
Comparing the SAM with the radiative transfer results reveals no strong difference. 
The ratio between the strength parameter of the two models $\log(a_{\rm RT} / a_{\rm SAM})$ 
is for the faint QSO $-0.1$ and for the bright one $0.1$. Considering the uncertainties in the
strength parameters, the 
two model yield identical results. We provide the ratios of the strength parameters between the
two models for all the studied environments and redshifts in Table \ref{tab:aParamTable}. 
The ratios show a strong scatter between the
different models. However due to the uncertainties in the strength parameter,
we attribute these differences to the inability of the analytical model to account for
the large scale density distribution.

\begin{figure}
\centering
\includegraphics[width=1.0\columnwidth]{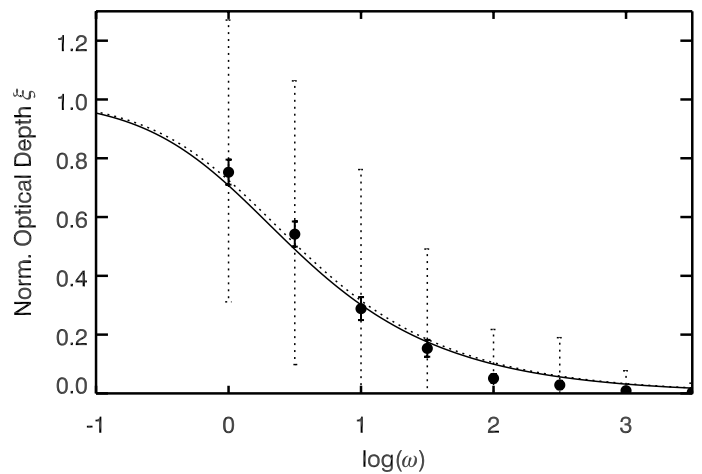}

\caption{\label{fig:randLOS} The combined proximity effect
signature of the SAM applied to 500 randomly selected lines of sight with random origin.
A $L_{\nu_{\rm LL}}=
10^{31}\textrm{ erg Hz}^{-1}\textrm{ s}^{-1}$ QSO is assumed. The continuous line
shows the fit of the analytical profile to the data.}
\end{figure}

To strengthen this conclusion, we have performed one additional test, where we applied
the SAM to 500 randomly selected lines of sight with random origins and directions
in the box and not just one specific origin. The resulting proximity effect profile is shown 
in Fig. \ref{fig:randLOS}.
The profile now smoothly follows the analytical model and does not show any fluctuations
around it. The strength parameter of $\log a = 0.11 \pm 0.10$ is as well consistent with
the SED corrected BDO model. The analytical model is thus a good description for a sample
of lines of sight with random origins and directions.

Thus the large shift in our simulations is solely due to the presence of large
scale overdensities around our sources. Further we cannot identify significant radiative 
transfer effect in addition to the 
one modelled with the SED correction term, since the SAM model gives similar results to the
radiative transfer simulations.

\begin{table}
\caption{\label{tab:aParamTable}Parameters of the $\log a$ fits to 
the combined proximity profile for the semi-analytical model (SAM) 
and the radiative transfer simulation (RT). We give $1\sigma$ errors.}

\tiny \centering

\begin{tabular*}{1.0\columnwidth}{@{\extracolsep{\fill}} cccccccc }
\hline 
\hline$z$ & $L_{\nu_{\rm LL}}^{\dagger}$ & 
\multicolumn{3}{c}{$\log (a_{\rm RT} / a_{\rm SAM})$} & 
\multicolumn{3}{c}{$\log a_{\rm RT}$}\\
 &  & halo & fil. & void & halo & filament & void\\
\hline
\noalign{\vskip\doublerulesep}
4.9 & $10^{31}$ & 0.1 & 0.1 & 0.2 & $0.4 \pm 0.1$ & $0.3 \pm 0.1$ & $0.3 \pm 0.1$\\
4.0 & $10^{31}$ & 0.3 & 0.0 & 0.3 & $0.6 \pm 0.1$ & $0.1 \pm 0.1$ & $0.6 \pm 0.1$\\
3.0 & $10^{31}$ & 0.3 & 0.0 & 0.1 & $0.7 \pm 0.1$ & $0.5 \pm 0.1$ & $0.4 \pm 0.2$\\
\hline
\noalign{\vskip\doublerulesep}
4.9 & $10^{32}$ & 0.2 & 0.1 & 0.1 & $0.2 \pm 0.1$ & $0.1 \pm 0.1$ & $0.2 \pm 0.1$\\
4.0 & $10^{32}$ & 0.4 & 0.0 & 0.7 & $0.2 \pm 0.2$ & $0.1 \pm 0.1$ & $0.7 \pm 0.2$\\
3.0 & $10^{32}$ & 0.5 & 0.1 & 0.1 & $0.3 \pm 0.3$ & $0.3 \pm 0.1$ & $0.3 \pm 0.1$\\
\hline
\end{tabular*}
\begin{list}{}{} 
\item[$\dagger$:] $\textrm{ erg Hz}^{-1}\textrm{ s}^{-1}$
\end{list}
\end{table}

\subsection{\label{sec:localEnv} Influence of the Host Environment} 

We present the resulting proximity effect profiles for the halo, filament, and void QSO host 
environments as a function of redshift in Fig. \ref{fig:ProxyMultiOmega}. The corresponding 
strength parameters are given in Table \ref{tab:aParamTable}. In the analysis the
unreliability region of $\log \omega \ge 2.4$ for the faint QSO and $\log \omega \ge 3.4$ 
for the bright QSO have been exculded. As we have seen in the 
previous Section, the local host environment of the QSO extends up to a comoving radius of 
$r \approx 2 \textrm{ Mpc } h^{-1}$. Hence, we expect the local environment to affect the 
proximity effect profiles up to $\log \omega \approx 1.5$ for the faint QSO and to
$\log \omega \approx 2.5$ for the bright one.

The effect of the local environment is significant in the halo case. Except for the 
$z=4.9$ bright QSO, the $\xi$-profile rises strongly when approaching the direct 
vicinity of the QSO. The $\xi$-values start to rise to such an extent, that they lie out of the 
plotted region and deviate strongly from the analytical model. 
In addition, the fluctuations in the normalised optical depth strongly
increase. This departure from the analytic profile turns out to pose a 
problem in the determination of the strength parameter. In order to obtain a reasonable fit,
the local environment needs to be excluded. 

For the filament and void environment however, such a departure is not seen as expected. 
Only in the profile of the $z=3$ faint QSO in the filament environment there is an enhancement 
at $\log \omega \approx 2.5$. In the filament and void environments, the scatter in the 
normalised optical depth decreases. The lack of density enhancements in the immediate
surroundings of the host thus reduces the scatter. The influence of the local environment also 
decreases with increasing luminosity, which is not only due to the reduced opacity, but as 
well due to the larger spatial coverage of each $\log \omega$ bin. 

Beyond the local environment's region, the $\xi$-values return 
to the smooth profile expected from the analytic model. At $z=4.9$ the different 
environments show identical profiles. With decreasing redshift however, the 
profiles in the different environments start to deviate from the analytical profile
without any clear trend. These 
deviations are also seen in the corresponding strength parameters. 
For example the faint halo and void case at $z=4$ coincide. Though
for the bright QSO, the strength parameters of the halo and the void do not agree anymore.
Now the halo and the filament coincide within the error bars. Due to the limitations of the
analytical model to account for the large scale overdensity present in the simulation data, 
the fitted strength parameters show large unsystematic fluctuations.
In Section \ref{sec:RTEffects} we have seen, that the local environment does not cause
deviation from the geometric dilution model beyond $2 \textrm{ Mpc h}^{-1}$.
The overionisation profile was not found to be influenced by the halo, filament, or void 
environment at large radii. Therefore we conclude that the local environment
does not influence the proximity effect profile at radii larger than the local
host environment. Furthermore the fluctuations in the 
strength parameter do not correlate with the source environment and solely 
arise from the fact, that the analytical model does not take the cosmological density distribution
into account.

\subsection{\label{sec:meanProfZ} Redshift Evolution}

Studying the strength parameters, no apparent evolution with redshift can be seen, 
if the fitting errors are taken into account. However the uncertainties of the strength parameter 
decrease with increasing redshift. Only in the faint halo case there is a hint of a 
decrease in proximity strength with increasing redshift. 

By looking at the $\xi$-profiles in Fig. \ref{fig:ProxyMultiOmega} a clear decrease in the
fluctuations around the mean profile can be seen with increasing redshift. This is due to the
fact, that the density fluctuations of the dark matter are not as developed yet at high redshifts
than at lower ones. A minor effect may result from the decreasing transmission of the Ly$\alpha$
forest at higher redshift, i.e. the narrow range of gas densities probed by the Ly$\alpha$ forest. The analysis 
of the large-scale overdensity as discussed in Section \ref{sec:largeScaleEnv} 
demonstrates however the dominance
of the evolution of the density field. There is a slight tendency of the profile to show a stronger shift at low 
redshifts than at $z=4.9$. This is especially prominent in the case of the 
$L_{\nu_{\rm LL}}=10^{32}\textrm{ erg Hz}^{-1}\textrm{ s}^{-1}$ QSO.

\subsection{\label{sec:meanProfLum} Luminosity Dependence}

The environmental bias of the proximity effect signature depends on the QSO luminosity.
For all our redshifts, the shift in the profile is smaller for the bright QSO than for the 
weak one. Also the variance of the signal is reduced with increasing QSO luminosity.
This is best seen for redshifts $z=4.9$ and $z=3$. 

The cause of the luminosity dependence lies in the fact that the $\omega$ scale is a function 
of the QSO luminosity. Therefore an increase in the QSO luminosity translates into larger
distances covered by each $\log \omega$ bin. This dampens the influence of the 
fluctuating density field and thus the variance in $\xi$. Hence luminous QSOs constrain
the proximity effect signature better.

\section{\label{sec:PESD}Proximity Effect Strength Distribution}

\begin{figure}
\includegraphics[width=1\columnwidth]{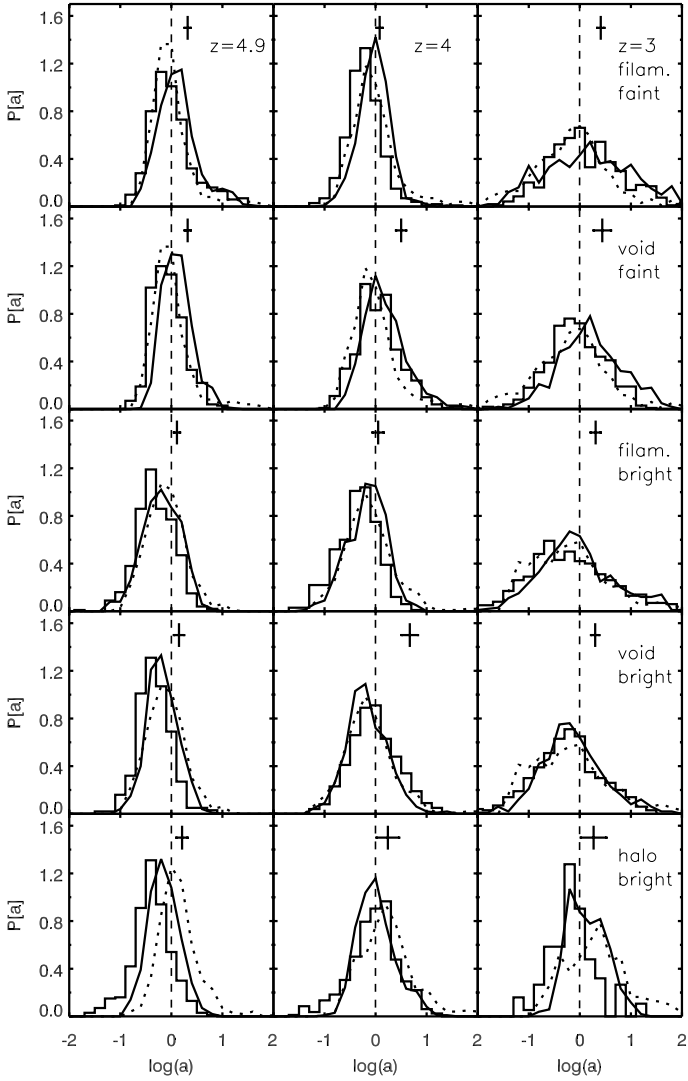}

\caption{\label{fig:aDistEnv}The distribution of the strength parameter 
$\log a$ of the proximity effect in the different lines of sight for the radiative 
transfer simulations (solid histogram), compared with RAM (dotted line)
and the SAM (solid line). The mean $a$-values from 
the radiative transfer simulations Table \ref{tab:aParamTable}) and their uncertainties 
are shown as vertical lines on top of each panel. The upper two panels show the low luminosity 
case. The lower three panels give results for the luminous QSO. From 
left to right the redshifts are 4.9, 4, and 3. The vertical dashed line marks $\log a = 0.0$.}

\end{figure}

Now we discuss our results on the distribution of the strength parameter determined on
individual lines of sight. The asymmetry of the $\xi$-distribution and the large scale
overdensity biases the mean
proximity effect profiles, as we have discussed in the previous Section. To 
avoid
this bias, \citet{DallAglio:2008ua} used the proximity strength distribution 
along individual lines 
of sight to measure the UVB flux, and proposed the peak of the distribution to 
provide 
a measure of the UVB. We test the proposed method with our 
simulations.

The proximity effect strength distribution is obtained by determining the 
$\xi$-profile for each 
individual line of sight and by fitting Eq. \ref{eq:omegaA} to the profile. In
contrast to the last Section we use all available $\xi$ data points (except the
ones inside the numerically unreliable region) in the
fitting procedure. Whenever the simulated data deviates strongly from the
analytical model due to the large variance of the 
$\xi$-values, it is not possible to determine a strength parameter. 
This problem is significant in the case of the faint QSO residing in a halo, 
where the proximity effect profile deviates from the analytical one at 
$\log \omega \ge 1.5$. There the fitting procedure failed in most cases and we
cannot derive a strength parameter distribution.

In Fig. \ref{fig:aDistEnv} we present the proximity effect strength 
distribution 
from our simulations. Only those lines of sight contribute to the distribution, 
for which a
strength parameter could be determined, however we always normalise the 
distributions $P(a)$ to unity. We employ the comparison models to investigate the 
various effects that alter the strength distribution. In order to
compare with the results obtained from the mean proximity effect profile, we 
marked 
the strength parameters of the mean profiles as given in Table 
\ref{tab:aParamTable}.

 All our models show a strength distribution that is clearly peaked. 
Furthermore the distribution reveals a large scatter of 
$\sigma(\log a) = 0.3,\, 0.4,\, 0.7$ at redshifts $z=4.9,\, 4,\, 3$. This scatter is
much larger than the variance of the $\log a$ values of the mean
analysis. All the models show a similar 
strength distribution as the respective ones of the RAM. In the case of the 
faint QSO, 
they all peak around 
$\log a = 0$. For the bright QSO, the peak is somewhat shifted to negative 
$\log a = -0.1$. This shift is as well present in the RAM. Therefore we
suspect that it is due to the small $\omega$ scale coverage for the bright QSO
and the corresponding uncertain fits. It follows that the peak of the 
strength distribution can be used to get unbiased estimates of the
proximity effect. 

Surprisingly the strength distribution is thus not sensitive 
to the large scale overdensity environment discussed in the mean analysis.
In the averaging process for deriving the mean profile, we are biased by 
the spherically averaged overdensities shown in Fig. \ref{fig:densityEnv}. 
By fitting the strength parameter individually, less weight is given to single 
$\log \omega$ bins deviating from the expected profile. This confirms the 
claim in \citet{DallAglio:2008ua} that using the proximity effect along
individual lines of sight leads to a stable unbiased estimate of the proximity 
effect and thus of the UVB. 

As in the mean analysis, we can investigate the influence of the local
environment, the quasar redshift, and the quasar luminosity. 
The shape of the strength parameter distribution is not influenced
by the QSO's local environment. Whether the QSO resides in a halo, a 
filament or a void, the maximum, median and the width of the 
distribution is not affected. Only for the faint quasar, the halo environment strongly influences
the proximity effect profile so that in many cases no reasonable fit could be given.  
The width of the distribution is found to increase strongly with redshift.
This is clearly a consequence of the increasing density inhomogeneities. 
Overall we find a large uncertainty of the proximity effect in individual 
sight lines. This is due to the limited $\omega$-scale coverage of the
simulated spectra, that are much smaller than in the observations analysed
in \citet{DallAglio:2008ua}.

\section{\label{sec:Conclusions}Conclusions}

We have performed high resolution Monte-Carlo radiative transfer simulations 
of the line of sight proximity effect. Using dark matter only simulations and a semi-analytic
model of the IGM, we applied our radiative transfer code in a post processing step. 
The aim of this paper was to identify radiative transfer influence on the 
proximity effect for QSOs with different Lyman limit luminosities, different 
host environments, and at different redshifts. 

Due to the low optical depths in the Ly$\alpha$ forest, we demonstrated that 
the on-the-spot approximation 
is insufficient in treating the proximity effect. 
Diffuse radiation from recombining electrons significantly contribute 
to the ionising photon budget over large distances and need to be taken into account 
in the simulations.

In the radiative transfer simulations we identified Lyman limit systems that cast shadows behind
them in the overionisation profile. Any Lyman limit system between the QSO and the 
observer will result in a suppression of the proximity effect. 
Behind such a system, the Ly$\alpha$ forest is not affected
by the additional QSO radiation and a detection of the proximity effect is not possible anymore.
However, Lyman limit systems appear only on rare lines of sight and 
influence only marginally the proximity effect statistics.

Mock spectra have been synthesized from the simulation results, and
similar methods as used in observations were applied to extract the observable
proximity effect signature. We compared these results with the widely
used assumption of the BDO model that only geometric dilution governs the 
over-ionisation region around the QSO. We have also used a random absorber
model applied extensively in the analysis of observational data. Furthermore a 
semi-analytical model was employed which analytically modelled the proximity effect signal 
onto spectra drawn from the UVB only simulations.

We have shown that differences in the shape of the UVB spectral energy
distribution function and the spectrum of the
QSO introduce a bias in the proximity effect strength. 
In the original formulation of the BDO model, the SED
of the two quantities was assumed to be equal. Differences in the SED of the QSO and the 
UVB result in a weakening or strengthening of the proximity effect. This bias can be 
analytically modelled through introducing a correction term in the BDO
formalism. We have tested the 
performance of the correction term. The correction term accounts mostly but not completely
for the effect. However the remaining difference lies
within the precision of the determination of the strength parameter. 

In all simulations and models, we have confirmed a clear proximity effect
profile. However, we found a strong scatter between different lines of sight. 
Cosmological density inhomogeneities introduced strong fluctuations in the normalised
optical depths of the proximity effect. They are caused by individual strong
absorbers along the line of sight. The influence is significant near the QSO and can
dominate over the standard proximity effect profile if the QSO resides in a 
massive halo. The fluctuations in the normalised optical depth are smaller, 
if more luminous QSOs are considered. There the real space coverage of each 
$\log \omega$ bin is larger and therefore the fluctuations decrease.
Additionally the fluctuations decrease with increasing redshift.

We have studied the influence of different QSO host environments on the proximity effect.
The QSO was placed in the most massive halo, in a random filament, and in a random
void. The local host environment of up to $r \approx 2 \textrm{ Mpc } h^{-1}$ is responsible
for deviations from the BDO model near the QSO. However it does not affect the proximity
effect at radii larger than $r \approx 2 \textrm{ Mpc } h^{-1}$, and the smooth proximity effect 
signature was regained. 

In the three environments selected for this study, random density enhancement
on scales up to $r \approx 15 \textrm{ Mpc } h^{-1}$ were present. These
overdensities introduced a bias in the mean analysis, that decreases the proximity strength. 
It is a strong effect, seen in averaging the profiles over the 500 lines of sight,
and then estimating the mean profile. This bias will affect the UVB 
measurement with this method. The influence of the bias decreases with 
increasing QSO luminosity.

Our radiative transfer results are complementary to the analysis of 
the proximity effect zone with a semi-analytical model by \citet{Faucher-Giguere:2008gf}. 
There the authors find a significant dependence of the 
combined proximity effect on the QSO host halo both by the action of the 
absorber clustering and the associated gas inflow. 
Coinciding with our analysis, a 
stronger proximity effect signal is found at lower redshift due to the higher gas density
inhomogeneities. 
The authors also discuss the nearly log-normal probability distribution of the 
flux decrement, and they find a strong dependence of the average overionisation 
in the near zone around the QSO host from the halo mass
on scales up to 1 Mpc proper radius.

The proximity effect strength distribution derived along single lines of sight does 
not show such a dependence on the  
large scale overdensities. The distributions were always consistent with the
random absorber model. 
This supports the analysis of real QSO spectra and the resulting unbiased 
estimate of the UV-background in \citet{DallAglio:2008ua}.
We found that the distributions are slightly shifted to negative $\log a$
values, since the $\omega$ range covered in the simulation was smaller 
for the high luminosity QSO.
The fluctuations in the normalised optical depth on single lines of sight are responsible for 
a large scatter in the derived strength distribution. The scatter increases
with decreasing redshift.
Since the $\omega$ scale in our analysis is limited, the fitted strength parameters show
large uncertainties. This contributes to the width of the strength parameter distribution. 
A complete coverage of the proximity effect region will reduce these
uncertainties on the strength distributions. 

To reliably study all the influences discussed here, the full proximity effect region needs to be 
simulated for which we have to make our radiative transfer code more efficient. It should also 
be noted, that Helium was not included in our simulations, but it could play an important role 
in additionally softening the QSO spectra. Helium would absorb a fraction of the hard flux 
produced by the QSO and could increase the influence of the QSO SED.

Finally we conclude that the QSO host environment, i.e. whether it sits in a
halo, a filament or a void, influences the proximity strength only
locally. Except shadowing by Lyman limit systems, the proximity effect does
not show radiative transfer influence other then the SED bias.

\appendix
\section{Strong absorber contamination of the normalised optical depth}

\begin{figure}
\includegraphics[width=1\columnwidth]{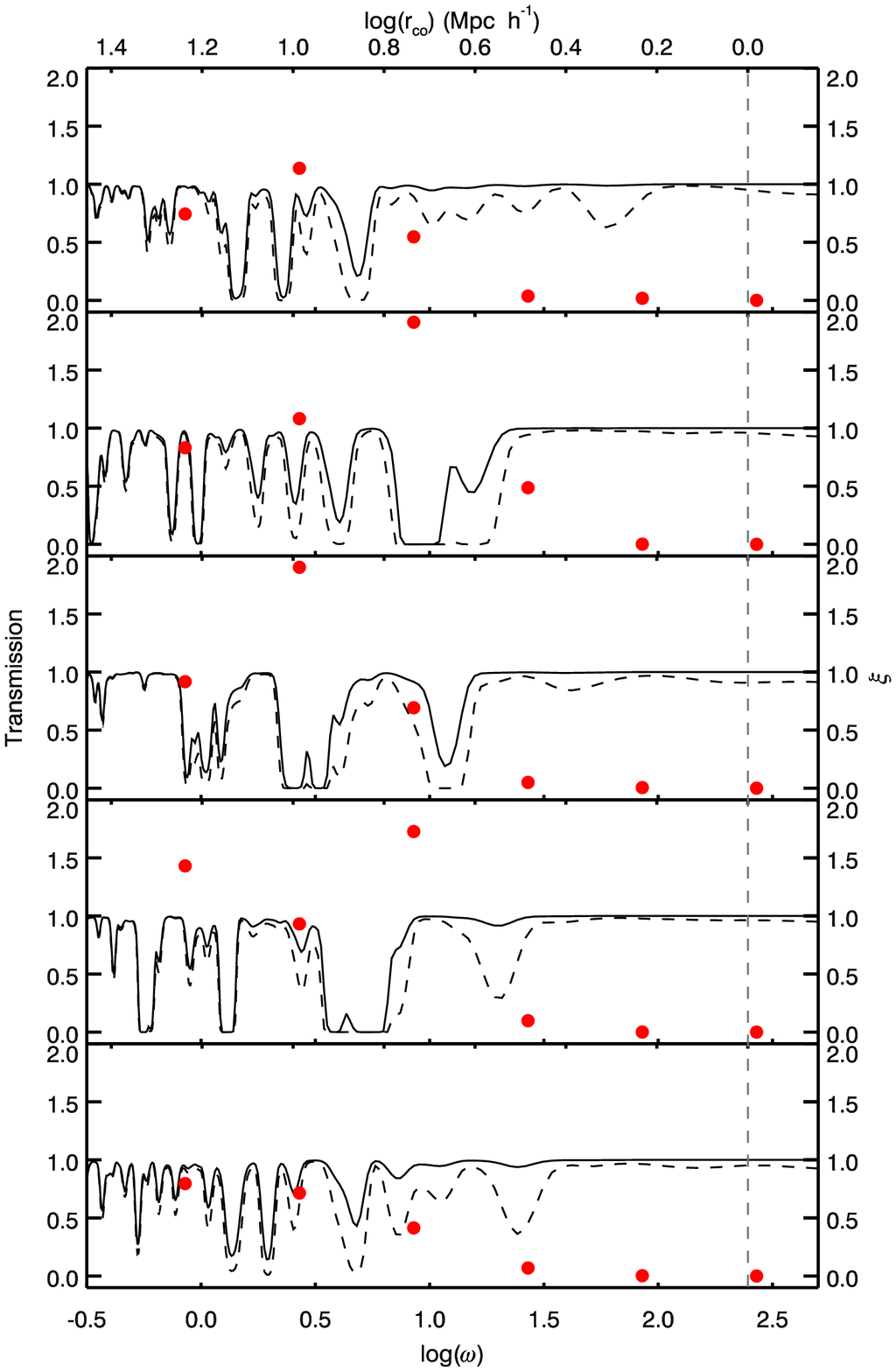}
\caption{\label{fig:appSpectra} Sample spectra taken from the $z=3$ low luminosity QSO
located in a void showing the radiative transfer simulations including the proximity effect (solid line)
and without it (dashed line). The spectra are given in $\log \omega$ scale and for better
visibility, no noise has been added to the spectra.
The normalised optical depths $\xi$ at each $\log \omega$ bin is overplotted by solid dots
to show the effect of absorption features on $\xi$.
The dashed vertical line marks the start of the unreliability region of 
$\log \omega \ge 2.4$.}
\end{figure}

In Section \ref{sec:meanProx} we have discussed the single line of sight variance
of the normalised optical depth $\xi$ and its dependency on luminosity and redshift. 
In this Appendix we illustrate where and how this variance arises. 

The normalised optical depth $\xi$ is the ratio of the effective optical depth
in the QSO spectra measured in $\log \omega$ bins over the effective optical depth
present in the unaffected Ly$\alpha$ forest. If the optical depth in a $\log \omega$ bin is larger
than in the unaffected forest, $\xi$ will be larger than unity. 
The range covered by the $\log \omega$ bins in the spectra decreases towards the QSO.
Therefore any strong absorption system present in the small ranges covered by high 
$\log \omega$ bins will bias the optical depth in the bins and therefore increase $\xi$ to values
above unity. This bias gives rise to the large variance in $\xi$ at large $\log \omega$. 
It is interesting to note, that an increase in the QSO Lyman luminosity by one dex also 
increases the covered proper distance in a $\log \omega$ bin by one dex. Thus the influence
of strong absorbers is reduced, the more luminous the QSO is.

In Fig. \ref{fig:appSpectra} we show five spectra taken from the $z=3$
QSO situated in a void with a luminosity of $L_{\nu_{\rm LL}}=
10^{31}\textrm{ erg Hz}^{-1}\textrm{ s}^{-1}$. The spectra are plotted in $\log \omega$ space
and the normalised optical depth measured in the $\log \omega$ bins is
overplotted by solid dots. The spectrum 
at the bottom panel shows an uncontaminated proximity effect profile. However
all the others panels illustrate the influence of high column density systems on the profile. 
In the second and third spectrum prominent absorption features elevate $\xi$ up to $\xi=2$.

\begin{figure*}
\begin{minipage}[t][1\totalheight]{0.49\textwidth}%
\includegraphics[width=1\columnwidth]{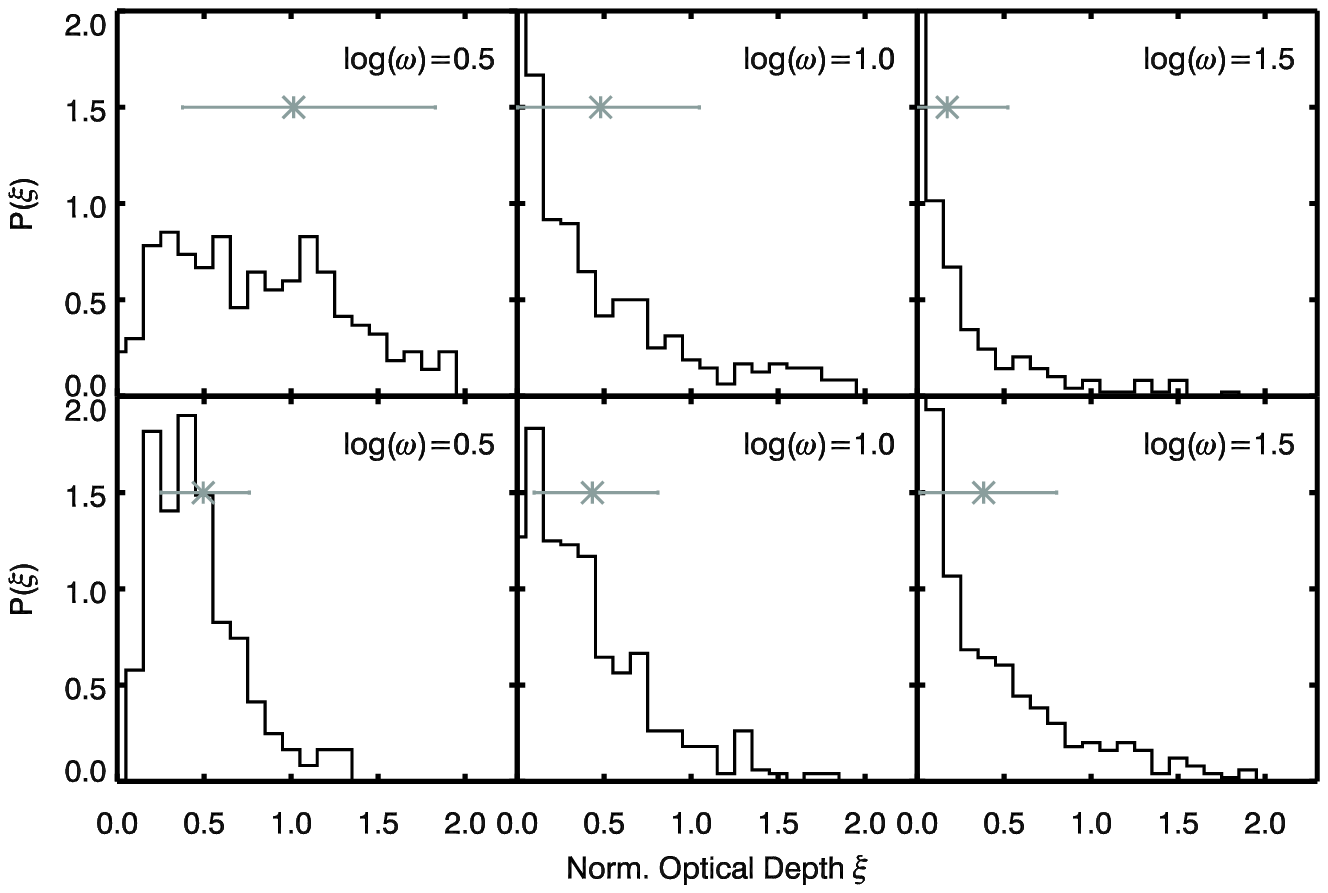}
\end{minipage}%
\hfill{}%
\begin{minipage}[t][1\totalheight]{0.49\textwidth}%
\includegraphics[width=1\columnwidth]{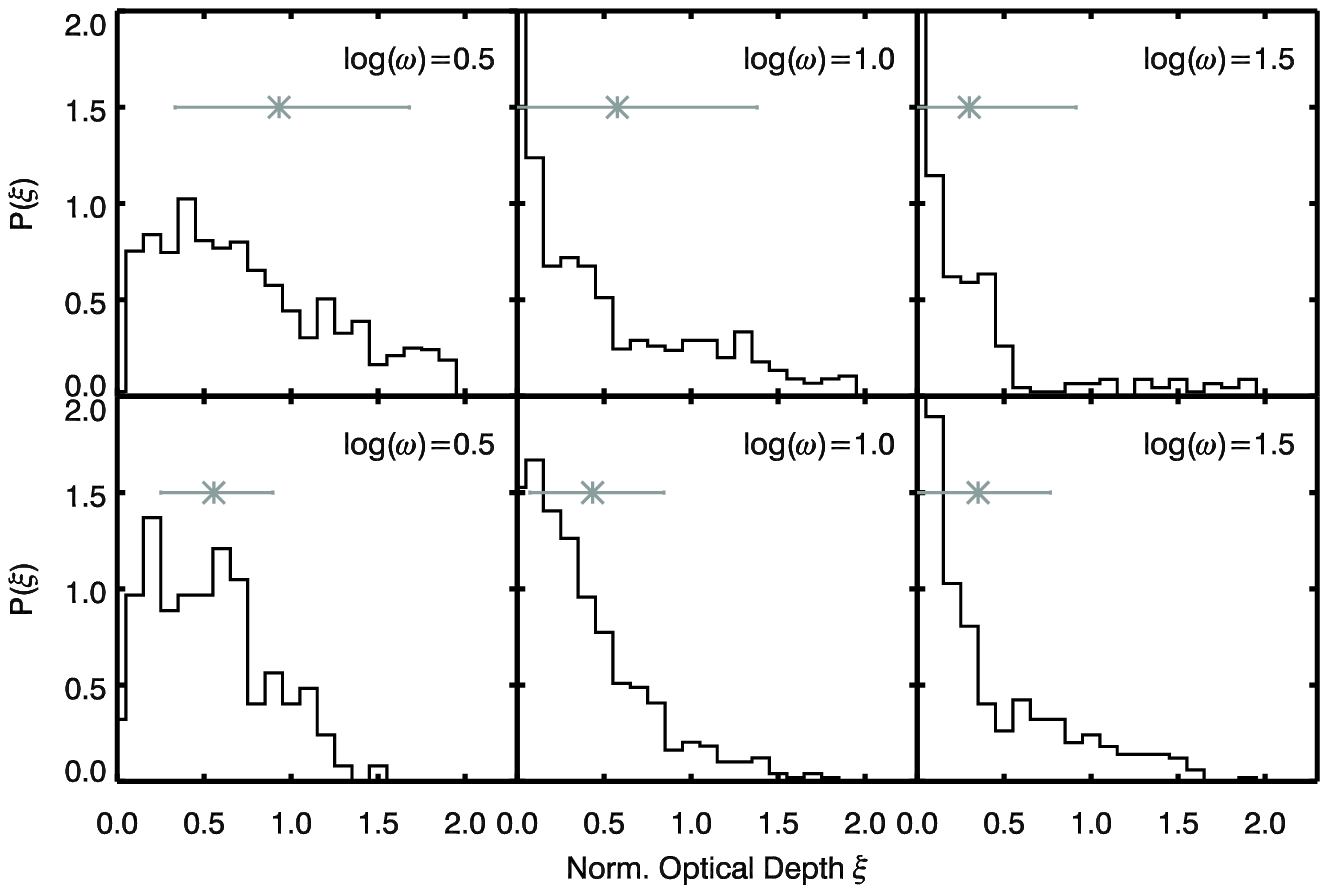}
\end{minipage}%
\caption{\label{fig:xiDistribution}Normalised optical depth $\xi$ distribution of the
void (left panel) and filament QSO (right panel) at redshift $z=3$ for three $\log \omega$
bins. The upper panels show results for the $L_{\nu_{\rm LL}}=
10^{31}\textrm{ erg Hz}^{-1}\textrm{ s}^{-1}$ QSOs, the lower one for the 
$L_{\nu_{\rm LL}}=
10^{32}\textrm{ erg Hz}^{-1}\textrm{ s}^{-1}$ QSO. The grey data points give the average
normalised optical depths with their variance as shown in Fig. \ref{fig:ProxyMultiOmega}.}
\end{figure*}

The distribution of $\xi$ in a $\log \omega$ bin arising from the influence of strong absorbers 
varies from bin to bin. In Fig. \ref{fig:xiDistribution} we show the $\xi$ distribution in three 
bins for QSOs located in a void and a filament at redshift $z=3$. Overplotted are the mean 
values and the corresponding standard deviations shown in Fig. 
\ref{fig:ProxyMultiOmega}. We note that for the bright QSO the $\log \omega = 0.5$ 
distribution is incomplete, since not all lines of sight reach the proper distance needed for 
this bin.

The distributions are anisotropic with an extended tail to large optical
depth. This behaviour reflects the probability distribution of the
simulated density field.  
The distribution of the high luminosity QSO is narrower than the one of the low luminosity QSO.
This is again due to the larger distances covered by the $\log \omega$ bins in the high 
luminous case. The width of the distribution covers a large $\xi$ range. 
The contamination of a single line of sight by strong absorption systems is 
thus an issue that can only be eliminated by analysing large number of QSO sight lines to 
reliably determine the mean proximity effect profile.

\begin{acknowledgements}
The authors thank Lutz Wisotzki for discussions and useful advice.
We are indebted to Tae-Sun Kim and Martin Haehnelt for constructive remarks.
We are grateful to the anonymous referee for the careful study and suggestions 
that helped to improve the paper.
The simulations have been carried out at the Konrad-Zuse-Zentrum f\"{u}r
Informationstechnik in Berlin, Germany using the AstroGrid-D. A.P.
thanks A. Maselli for helpful discussions about CRASH and this study.
Further A.P. acknowledges support in parts by the German Ministry
for Education and Research (BMBF) under grant FKZ 05 AC7BAA. A.D.A
acknowledges support by the Deutsche Forschungsgemeinschaft under
Wi 1369/21-1. 
\end{acknowledgements}

\bibliography{biblio}

\end{document}